\documentclass[twocolumn,modern,astrosymb,tighten,trackchanges,article]{aastex631}
\usepackage{comment}
\usepackage{CJK}
\usepackage{amsmath}
\usepackage{appendix}
\usepackage{hyperref}

\newcommand{\ie}{i.e.\/}
\newcommand{\eg}{e.g.\/}

\submitjournal{The Astrophysical Journal}
\accepted{Sept. 12, 2022}

\shorttitle{HCN and HNC as an Interstellar Heating Probe}
\shortauthors{Behrens et al.}

\graphicspath{{./}{figures/}}
\begin{document}
\begin{CJK*}{UTF8}{gbsn}

\title{Tracing Interstellar Heating: An ALCHEMI Measurement of the HCN Isomers in NGC\,253}

\author[0000-0002-2333-5474]{Erica Behrens}
\altaffiliation{Student at the National Radio Astronomy Observatory}
\affiliation{Department of Astronomy, University of Virginia, P.~O.~Box 400325, 530 McCormick Road, Charlottesville, VA 22904-4325}
\author[0000-0003-1183-9293]{Jeffrey G.~Mangum}
\affiliation{National Radio Astronomy Observatory, 520 Edgemont Road,
  Charlottesville, VA  22903-2475, USA}
%\email{jmangum@nrao.edu}
%
\author[0000-0003-4025-1552]{Jonathan Holdship}
\affiliation{Leiden Observatory, Leiden University, PO Box 9513, NL-2300 RA Leiden, The Netherlands}
\affiliation{Department of Physics and Astronomy, University College London, Gower Street, London WC1E 6BT}
\author[0000-0001-8504-8844]{Serena Viti}
\affiliation{Leiden Observatory, Leiden University, PO Box 9513, NL-2300 RA Leiden, The Netherlands}
\affiliation{Department of Physics and Astronomy, University College London, Gower Street, London WC1E 6BT}
\author[0000-0002-6824-6627]{Nanase Harada}
\affiliation{National Astronomical Observatory of Japan, 2-21-1 Osawa, Mitaka, Tokyo 181-8588, Japan}
\affiliation{Department of Astronomy, School of Science, The Graduate University for Advanced Studies (SOKENDAI), 2-21-1 Osawa, Mitaka, Tokyo, 181-1855 Japan}

\author[0000-0001-9281-2919]{Sergio Mart\'in}
\affiliation{European Southern Observatory, Alonso de C\'ordova, 3107, Vitacura, Santiago 763-0355, Chile}
\affiliation{Joint ALMA Observatory, Alonso de C\'ordova, 3107, Vitacura, Santiago 763-0355, Chile}
%\email{Sergio.Martin@eso.org}

\author[0000-0001-5187-2288]{Kazushi Sakamoto}
\affiliation{Institute of Astronomy and Astrophysics, Academia Sinica, 11F of AS/NTU
Astronomy-Mathematics Building, No.1, Sec. 4, Roosevelt Rd, Taipei 10617, Taiwan}
%\email{ksakamoto@asiaa.sinica.edu.tw}

\author[0000-0002-9931-1313]{Sebastien Muller}
\affiliation{Department of Space, Earth and Environment, Chalmers University of Technology, Onsala Space Observatory, SE-43992 Onsala, Sweden}
%\email{mullers@chalmers.se}

\author[0000-0001-8153-1986]{Kunihiko Tanaka}
\affil{Department of Physics, Faculty of Science and Technology, Keio University, 3-14-1 Hiyoshi, Yokohama, Kanagawa 223--8522 Japan}
%\email{ktanaka@phys.keio.ac.jp}

\author[0000-0002-6939-0372]{Kouichiro Nakanishi}
\affiliation{National Astronomical Observatory of Japan, 2-21-1 Osawa, Mitaka, Tokyo 181-8588, Japan}
\affiliation{Department of Astronomy, School of Science, The Graduate University for Advanced Studies (SOKENDAI), 2-21-1 Osawa, Mitaka, Tokyo, 181-1855 Japan}
%\email{nakanisi.k@nao.ac.jp}

\author[0000-0002-7758-8717]{Rub\'en~Herrero-Illana}
\affiliation{European Southern Observatory, Alonso de C\'ordova, 3107, Vitacura, Santiago 763-0355, Chile}
\affiliation{Institute of Space Sciences (ICE, CSIC), Campus UAB, Carrer de Magrans, E-08193 Barcelona, Spain}
%\email{rherrero@eso.org}

\author{Yuki Yoshimura}
\affiliation{Institute of Astronomy, Graduate School of Science,
The University of Tokyo, 2-21-1 Osawa, Mitaka, Tokyo 181-0015, Japan}
%\email{yyoshimura@ioa.s.u-tokyo.ac.jp}

%\author[0000-0002-4355-6485]{Stefanie~M\"uhle}
%\affiliation{Argelander-Institut f\"ur Astronomie, Universit\"at Bonn, Auf dem H\"ugel 71, D-53121 Bonn, Germany}(*)
%\email{muehle@astro.uni-bonn.de}

% Other authors.  These authors only appear on the full author list which
% appears on the last page of the article.  Order is alphabetical.

% \author[0000-0002-5828-7660]{Susanne Aalto}
% \affiliation{Department of Space, Earth, and Environment, Onsala Space Observatory, Chalmers University of Technology, Onsala Observatory, SE-439 92 Onsala, Sweden}
% %\email{saalto@chalmers.se}

\author[0000-0002-1316-1343]{Rebeca Aladro}
\affiliation{Max-Planck-Institut f\"ur Radioastronomie, Auf dem H\"ugel 69, 53121 Bonn, Germany}
% %\email{aladro@mpifr-bonn.mpg.de}

 \author[0000-0001-8064-6394]{Laura Colzi}
\affiliation{Centro de Astrobiolog\'ia (CSIC-INTA), Ctra. de Ajalvir Km. 4, 28850, Torrej\'on de Ardoz, Madrid, Spain}
%\affiliation{INAF-Osservatorio Astrofisico di Arcetri, Largo E. Fermi 5, I-50125, Florence, Italy 5}
% %\email{lcolzi.astro@gmail.com}

% \author[0000-0001-8509-1818]{Gary A.~Fuller}
% \affiliation{Jodrell Bank Centre for Astrophysics, Department of Physics \& Astronomy, School of Natural Sciences, The University of Manchester, Oxford Road, Manchester, M13 9PL, UK}
% %\email{g.fuller@manchester.ac.uk}

% \author[0000-0003-0444-6897]{Santiago Garc\'{\i}a-Burillo}
% \affiliation{Observatorio Astron\'omico  Nacional (OAN-IGN), Observatorio de Madrid, Alfonso XII, 3, 28014-Madrid, Spain}
% %\email{s.gburillo@oan.es}

% \author[0000-0001-6431-9633]{Adam G.~Ginsburg}
% \affiliation{Department of Astronomy, University of Florida, Bryant Space Science Center, Stadium Road, Gainesville, FL 32611, USA}
% %\email{adamginsburg@ufl.edu}

% \author[0000-0002-2554-1837]{Thomas R.~Greve}
% \affiliation{Dept. of Physics and Astronomy, University College London, Gower Street, London WC1E6BT, UK}
% \affiliation{Cosmic Dawn Center}
% %\email{t.greve@ucl.ac.uk}

\author[0000-0001-6527-6954]{Kimberly L. Emig}
\altaffiliation{Jansky Fellow of the National Radio Astronomy Observatory}
\affiliation{National Radio Astronomy Observatory, 520 Edgemont Road,
  Charlottesville, VA  22903-2475, USA}

 \author[0000-0002-7495-4005]{Christian Henkel}
 \affiliation{Max-Planck-Institut f\"ur Radioastronomie, Auf dem H\"ugel   69, 53121 Bonn, Germany}
 \affiliation{Astronomy Department, Faculty of Science, King Abdulaziz
   University, P.~O.~Box 80203, Jeddah 21589, Saudi Arabia}
% %\email{chenkel@mpifr-bonn.mpg.de}

\author[0000-0002-1227-8435]{Ko-Yun Huang}
\affiliation{Leiden Observatory, Leiden University, PO Box 9513, NL-2300 RA Leiden, The Netherlands}

\author[0000-0003-3537-4849]{P. K. Humire}
\affiliation{Max-Planck-Institut f\"ur Radioastronomie, Auf dem H\"ugel 69, 53121 Bonn, Germany}
% %\email{phumire@mpifr-bonn.mpg.de} 

% \author[0000-0001-9162-2371]{Leslie K.~Hunt}
% \affiliation{INAF-Osservatorio Astrofisico di Arcetri, Largo E. Fermi 5, I-50125 Firenze, Italy}
% %\email{hunt@arcetri.astro.it}

% \author[0000-0001-9452-0813]{Takuma Izumi} 
% \affiliation{National Astronomical Observatory of Japan, 2-21-1 Osawa, Mitaka, Tokyo 181-8588, Japan} 
% %\email{takuma.izumi@nao.ac.jp} 

% \author[0000-0002-4052-2394]{Kotaro Kohno (河野孝太郎)}
% \affiliation{Institute of Astronomy, Graduate School of Science,
% The University of Tokyo, 2-21-1 Osawa, Mitaka, Tokyo 181-0015, Japan}
% \affiliation{Research Center for the Early Universe, Graduate
% School of Science, The University of Tokyo, 7-3-1 Hongo, Bunkyo-ku,
% Tokyo 113-0033, Japan}
% %\email{kkohno@ioa.s.u-tokyo.ac.jp}

\author[0000-0001-9436-9471]{David S.~Meier}
\affiliation{New Mexico Institute of Mining and Technology, 801 Leroy Place, Socorro, NM 87801, USA}
\affiliation{National Radio Astronomy Observatory, PO Box O, 1003 Lopezville Road, Socorro, NM 87801, USA}
% %\email{david.meier@nmt.edu}

% \author{Taku Nakajima}
% \affiliation{Institute for Space-Earth Environmental Research, Nagoya University, Furo-cho, Chikusa-ku, Nagoya, Aichi 464-8601, Japan}
% %\email{nakajima@isee.nagoya-u.ac.jp}

% \author[0000-0003-0563-067X]{Yuri Nishimura}
% \affiliation{Institute of Astronomy, The University of Tokyo, 
% 2-21-1, Osawa, Mitaka, Tokyo 181-0015, Japan}
% \affiliation{ALMA Project, National Astronomical Observatory of Japan, 
% 2-21-1, Osawa, Mitaka, Tokyo 181-8588, Japan}
%% %\email{yuri@ioa.s.u-tokyo.ac.jp}

 \author[0000-0002-2887-5859]{V\'ictor M.~Rivilla}
 \affiliation{Centro de Astrobiolog\'ia (CSIC-INTA), Ctra. de Ajalvir Km. 4, 28850, Torrej\'on de Ardoz, Madrid, Spain}
% \affiliation{INAF-Osservatorio Astrofisico di Arcetri, Largo E. Fermi 5, I-50125, Florence, Italy 5}
% %\email{vmrivilla@gmail.com}

% \author[0000-0003-1946-8482]{Mamiko T.~Sato}
% \affiliation{Department of Space, Earth and Environment, Chalmers University of Technology, Onsala Observatory, SE-439 92 Onsala, Sweden}
% %\email{mamiko@chalmers.se}

% \author[0000-0001-6788-7230]{Shuro Takano}
% \affiliation{Department of Physics, General Studies,
% College of Engineering, Nihon University, Tamura-machi,
% Koriyama, Fukushima 963-8642, Japan}
% %\email{takano.shuro@nihon-u.ac.jp}

 \author[0000-0001-5434-5942]{Paul P.~van der Werf}
 \affiliation{Leiden Observatory, Leiden University,
     PO Box 9513, NL - 2300 RA Leiden, The Netherlands}
 %\email{pvdwerf@strw.leidenuniv.nl}
 
% \author[0000-0002-8809-2725]{Yongxiong Wang (王永雄)}
% %\author[0000-0002-8809-2725]{Yongxiong Wang}
% \affiliation{Jodrell Bank Centre for Astrophysics, Department of Physics
% \& Astronomy, School of Natural Sciences, The University of Manchester,
% Oxford Road, Manchester, M13 9PL, UK}
% %\email{yongxiong.wang@postgrad.manchester.ac.uk}

\collaboration{20}{(ALMA Comprehensive High-resolution Extragalactic Molecular Inventory (ALCHEMI) collaboration)} 

\correspondingauthor{Erica Behrens} \email{eb7he@virginia.edu}

% Use \input to insert email list here
%\input{emaillist.tex}  

%%%%%%%%%%%%%%%%%%%%%%%%%%%%%%%%%%%%%%%%%%%%%%%%%%%%%%%%

\begin{abstract}

We analyze HCN and HNC emission in the nearby starburst galaxy NGC\,253 to investigate its effectiveness in tracing heating processes associated with star formation. This study uses multiple HCN and HNC rotational transitions observed using ALMA via the ALCHEMI Large Program. To understand the conditions and associated heating mechanisms within NGC\,253's dense gas, we employ Bayesian nested sampling techniques applied to chemical and radiative transfer models which are constrained using our HCN and HNC measurements. We find that the volume density $n_{\text{H}_{2}}$ and cosmic ray ionization rate (CRIR) $\zeta$ are enhanced by about an order of magnitude in the galaxy's central regions as compared to those further from the nucleus. In NGC\,253's central GMCs, where observed HCN/HNC abundance ratios are lowest, $n \sim 10^{5.5}$ cm$^{-3}$ and $\zeta \sim 10^{-12}$ s$^{-1}$ (greater than $10^4$ times the average Galactic rate). We find a positive correlation in the association of both density and CRIR with the number of star formation-related heating sources (supernova remnants, HII regions, and super hot cores) located in each GMC, as well as a correlation between CRIRs and supernova rates. Additionally, we see an anticorrelation between the HCN/HNC ratio and CRIR, indicating that this ratio will be lower in regions where $\zeta$ is higher. Though previous studies suggested HCN and HNC may reveal strong mechanical heating processes in NGC\,253's CMZ, we find cosmic ray heating dominates the heating budget, and mechanical heating does not play a significant role in the HCN and HNC chemistry.

\end{abstract}

%%%%%%%%%%%%%%%%%%%%%%%%%%%%%%%%%%%%%%%%%%%%%%%%%%%%%%%%

\section{Introduction} \label{sec:intro}

\begin{figure*}
    \centering
    \includegraphics[trim = 0mm 0mm 5mm 0mm, scale=0.68]{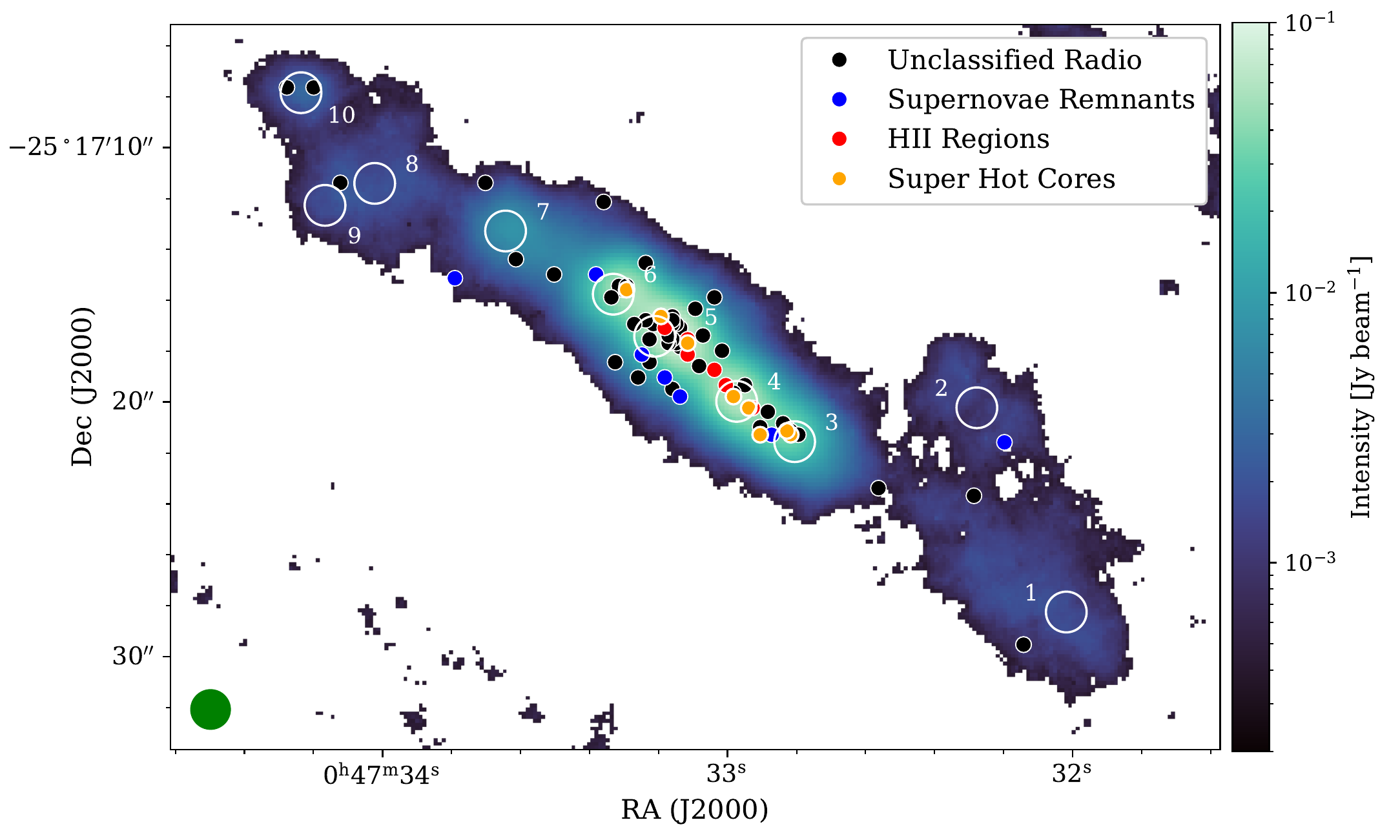}
    \caption{Location of radio continuum sources \citep{ua97} and super hot cores \citep{RV2020} within the NGC\,253 CMZ plotted over the 212 GHz ALCHEMI dust continuum emission. Numbered white circles indicate GMCs identified in \cite{leroy15}. The beam size of 1.6\,arcseconds is shown by the green circle in the bottom left corner.}
    \label{fig:ua_srcs}
\end{figure*}

Starburst galaxies have long been subjects of interest in astrophysical research due to their extreme star-forming environments as compared to the Milky Way. Observing starburst galaxies allows us to study how stars form in regions with higher densities, temperatures, and velocity dispersion. However, the physical conditions in extragalactic star-forming regions are not well understood due to limitations in resolving substructure and thus examining conditions on giant molecular cloud scales at mm and sub-mm wavelengths \citep{leroy18}. Many processes associated with star formation (mechanical heating in the form of shocks and turbulence from supernova explosions, radiative heating from massive stars, ionization by cosmic rays from supernova remnants, etc.) have competing effects on the interstellar medium (ISM).  Determining the influence of each of these physical processes on extragalactic star-forming regions is crucial to our understanding of the chemical and physical processes that guide star formation in starburst environments.

We study the nearby galaxy NGC\,253 as a laboratory for exploring how the current generation of stars affects future star formation in a starburst galaxy. It has an inclination of 76$^{\circ}$ \citep{incl_McCormick}, and at a distance of $3.5\pm0.2$\,Mpc \citep{dist_Rekola}, NGC\,253 is an ideal target for studying extragalactic star formation. NGC\,253 features a Central Molecular Zone (CMZ) spanning $\sim$800\,pc across which hosts at least 10 Giant Molecular Clouds (GMCs) identified via the dense gas tracers HCN, HCO$^{+}$, and CS \citep[][Appendix~\ref{sec:GMCpos} and Figure~\ref{fig:ua_srcs}]{leroy15}. Despite hosting a star formation rate of 5 M$_{\odot}$ yr$^{-1}$ across the entire galaxy, the central kiloparsec accounts for 40\% of that rate, forming stars at a rate of 2 M$_{\odot}$ yr$^{-1}$. This centrally-concentrated star formation results in NGC\,253's classification as a nuclear starburst \citep{leroy15}.

To capitalize on NGC\,253's ideal positioning and chemical complexity \citep{Aladro15, Martin19}, the ALMA Comprehensive High Resolution Molecular Inventory (ALCHEMI) observing program was conducted.  ALCHEMI is an ALMA large program which imaged the NGC\,253 CMZ over a frequency range of 84.2 to 373.2 GHz \citep{ALCHEMI-ACA}. ALCHEMI has cultivated the investigation of the rich chemical environment within the NGC\,253 CMZ using a comprehensive molecular inventory to trace chemical and physical processes associated with starburst environments. ALCHEMI allows for the study of GMC-scale structures ($\sim50$ pc) located in NGC\,253's CMZ due to its sensitivity to physical size scales from 255\,pc ($15^{\prime\prime}$) to 28\,pc ($1.^{\prime\prime}6$). 

This paper is one in a series of ALCHEMI projects that analyzes the conditions in the NGC\,253 CMZ using molecular signatures \citep{ALCHEMI-ACA,harada21,holdship_c2h,Haasler2022,Holdship2022,Humire2022}. Additionally, in this paper we explore how molecular emission can trace heating processes associated with star formation in this active environment.

\begin{figure*}
    \centering
    \includegraphics[scale=0.87]{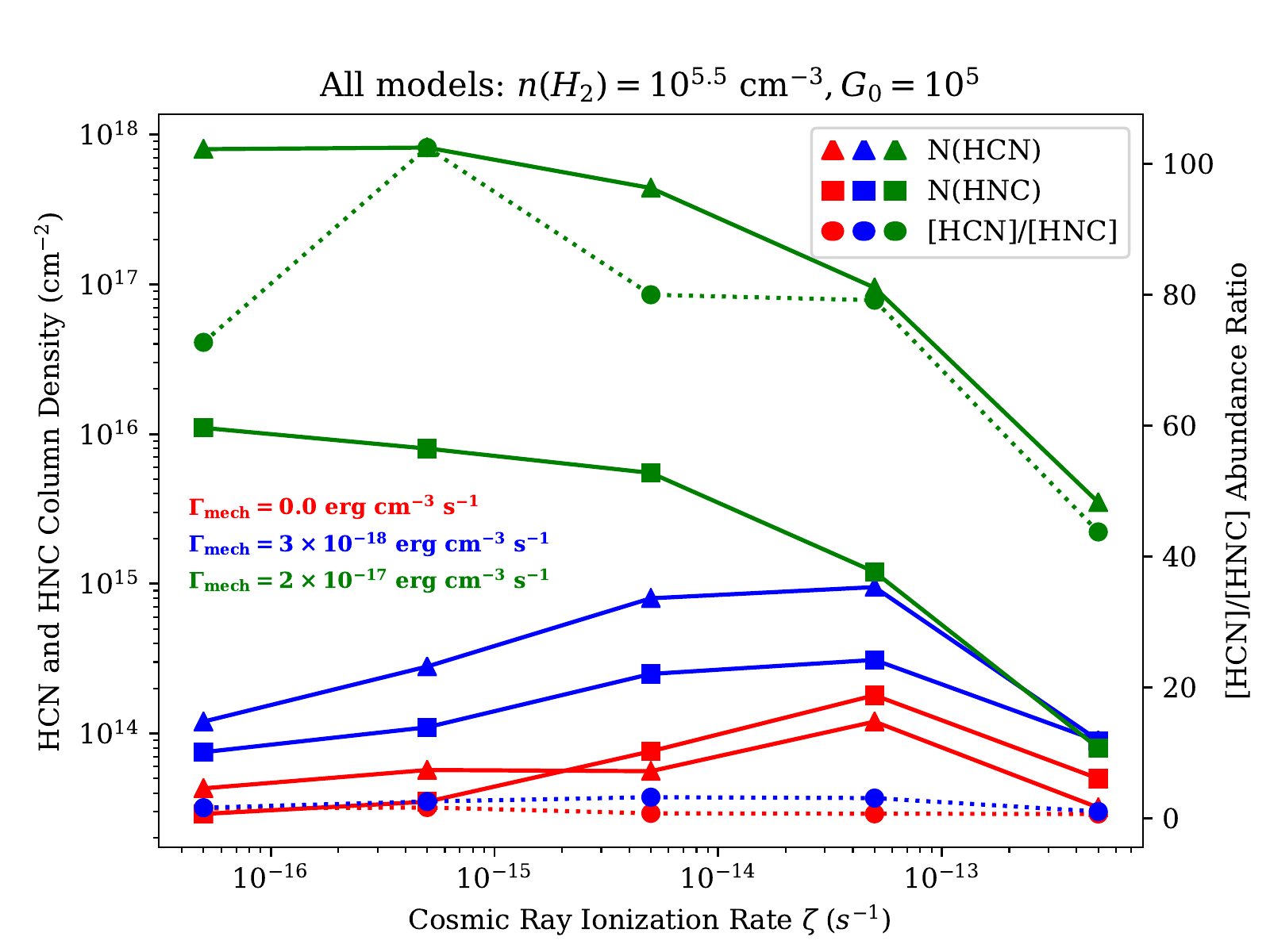}
    \caption{HCN and HNC column density (left axis) and column density ratio (right axis) as a function of CR ionization rate $\zeta$ and mechanical heating rate $\Gamma_{mech}$ from the PDR models presented by \cite{meijerink11}.  Volume density $n_{\text{H}_2}$ and far-UV (FUV) radiation field intensity G$_0$ are fixed at $10^{5.5}$\,cm$^{-3}$ and $10^5$\,Habing, respectively, in these models.}
    \label{fig:meijerink_ratio}
\end{figure*}

The strong star formation activity in NGC\,253 is evidenced by at least 64 individual compact radio continuum sources within the CMZ \citep{ua97}, particularly concentrated in GMCs 3--6 (Figure \ref{fig:ua_srcs}). \cite{ua97} measure spectral indices $\alpha$ ($S_\nu \propto \nu^\alpha$) for 23 of these sources using wavelengths ranging from 1.3 to 20 cm with resolutions between 1 and 15 pc. Of these 23 spectral index measurements, 17 have spectral index uncertainties $\sigma_{\alpha}$ of less than 0.4. About half of the sources in this subset are believed to be supernova remnants due to a measured spectral index below $-0.4$, which is indicative of synchrotron radiation. The remaining usable sources with $\sigma_{\alpha} < 0.4$ have spectral indices $\alpha$ ranging from 0.0 to 0.2, which is consistent with free-free emission from HII regions. \cite{ua97} note that the majority of sources emitting free-free radiation lie along the galaxy disk major axis, whereas the synchrotron sources lie farther away from the midline. The brightest of these radio sources \citep[TH2,][]{th85} is located in GMC 5 and associated with the nucleus of the galaxy, within 1$^{\prime\prime}$ of the galaxy's kinematic center \citep{ms10}. Other sources associated with star formation in the CMZ are proto-Super Star Clusters \citep{leroy18} containing Super Hot Cores (SHCs) identified by \cite{RV2020}\footnote{Note that the measurements identifying super hot cores sample only the part of the NGC 253 CMZ encompassing GMCs 3 through 6.} using vibrationally-excited HC$_{3}$N emission. These measurements suggest that the NGC\,253 CMZ GMCs are currently at different stages of evolution. 

\section{HCN and HNC in Galaxies} \label{sec:gal_hcn}

To investigate the physical conditions in the NGC\,253 CMZ, we can use combinations of chemical tracers from ALCHEMI's robust dataset that highlight the mechanisms involved in star formation and its effects on the environment. The combination that we will explore in this article is HCN and its isomer HNC.  HCN and HNC have similar energy level structures and dipole moments (differ by 2.2\%); hence their abundance ratio is often used as a probe of gas chemical conditions \citep[\eg][]{Goldsmith1986,Schilke1992,Herbst2000}. 

Additionally, HCN and HNC transitions are relatively bright in an extragalactic context and thus easy to detect. Studies of the HCN and HNC emission have been reported toward a wide range of galaxy types, including normal, luminous infrared, and active galactic nucleus-dominated galaxies \citep{Aalto2002A&A,Aalto2007A&A,Aalto2007aA&A,Aalto2012A&A,Costagliola2011A&A,Costagliola2015A&A,Green2016MNRAS,Greve2009ApJ,Imanishi2013AJ,Kamenetzky2011ApJ,Li2021MNRAS,Perez-Beaupuits2007A&A}, as well as high-redshift galaxies \citep{Spilker2014ApJ}.  Using HCN and HNC transitions ranging from J=$1-0$ to $4-3$ the HCN/HNC spectral line integrated intensity ratio ranges from $\sim 1 - 5$.  In a few luminous infrared galaxies the HCN/HNC spectral line intensity ratio is measured to be less than 1 \citep{Aalto2007A&A}.  In these galaxies a model which includes infrared excitation of the lowest-energy vibrational bending mode is used to explain this unusual HCN/HNC ratio. 

In our own Galaxy, the HCN/HNC ratio is very close to unity across different environments, from dense quiescent molecular clouds to star-forming regions \citep[\eg][]{irvine84,hirota98}. Within the low-A$_v$ and high-UV flux environments found in Planetary Nebulae, HNC is more readily destroyed due to the warming of the environment from UV radiation \citep{Bublitz2022A&A}. However, in high-A$_v$ regions where high-mass star formation dominates, the HCN/HNC abundance ratio has been found to be much higher \citep{Schilke1992}. This is believed to be due to the destruction of HNC (rather than an enhancement of HCN) via an isomerization reaction which occurs at relatively high temperatures. However, the temperature barrier for this reaction is uncertain. Theoretical studies suggest a barrier of 1200\,K, while observational results are better explained by a 200\,K barrier \citep{gran14, hacar20}. Despite these conflicting results, it would be expected that at high temperatures, the abundance of HCN would increase with respect to HNC. 

Previous studies have used ratios of formaldehyde transitions to derive kinetic temperatures $T_{\rm{K}}$ in NGC\,253's central GMCs (3--7), finding that $T_K \gtrsim 50$ K on 5$^{\prime\prime}$ ($\sim$80 pc) scales and $T_{K} \gtrsim 300$ K on $\lesssim1^{\prime\prime}$ ($\lesssim16$ pc) scales \citep[\eg][]{mangum19}. It is unclear exactly which mechanisms are raising the kinetic temperatures to this level, but one possible explanation is mechanical heating as a result of shocks generated by supernova explosions and cloud-cloud collisions, as well as outflows from young stars \citep{Mauersberger2003}. \cite{meijerink11} suggest that mechanical heating consistent with the star formation activity in starburst galaxies could raise temperatures to over 100 K and up to 1000 K in regions of lower column density ($\lesssim5 \times 10^{21}$ cm$^{-2}$) for volume densities of 10$^{5.5}$ cm$^{-3}$. \cite{meijerink11} also find that mechanical heating that would raise the kinetic temperature to such values could increase the HCN/HNC abundance ratio by up to two orders of magnitude compared to its Milky Way value in quiescent clouds, suggesting that this ratio could be a good mechanical heating indicator (Figure~\ref{fig:meijerink_ratio}). \cite{kaz12} echo these results. \cite{hacar20} propose using the HCN/HNC abundance ratio as a kinetic temperature probe. 

Alternatively, cosmic rays, without the addition of mechanical heating, could be responsible for the kinetic temperatures measured in NGC\,253's CMZ \citep{bayet11,papadopoulos}. However, high rates of cosmic ray ionization may depress the HCN/HNC abundance ratio, as suggested by the analyses presented in \cite{bayet11} and \citet[][Figure~\ref{fig:meijerink_ratio}]{meijerink11}, which predict HCN and HNC abundances as a function of cosmic ray ionization rate. It is important to note, however, that these studies used models that couple temperature and chemical abundance calculations, where cosmic ray ionization rate affects the temperature. Thus, the effect of cosmic ray ionization and cosmic ray heating of the gas are difficult to separate. In order to fully differentiate between the contributions of cosmic ray chemistry and cosmic ray heating on the molecular ISM, we treat cosmic rays and heating separately in our models.

Our work combines ALCHEMI observations with chemical and physical modeling in order to ascertain the mechanisms driving the high kinetic temperatures in the nucleus of NGC\,253. In Section \ref{sec:data} we describe our ALCHEMI HCN and HNC isomer observations. We present the methods and results of our chemical modeling analysis in Section \ref{sec:model_methods}. Section \ref{sec:disc} discusses the implications of our combined observational and modeling results, and we summarize our findings in Section \ref{sec:conc}.

\begin{figure*}[htbp]
\includegraphics[trim = 5mm 7mm 5mm 5mm, scale=0.73]{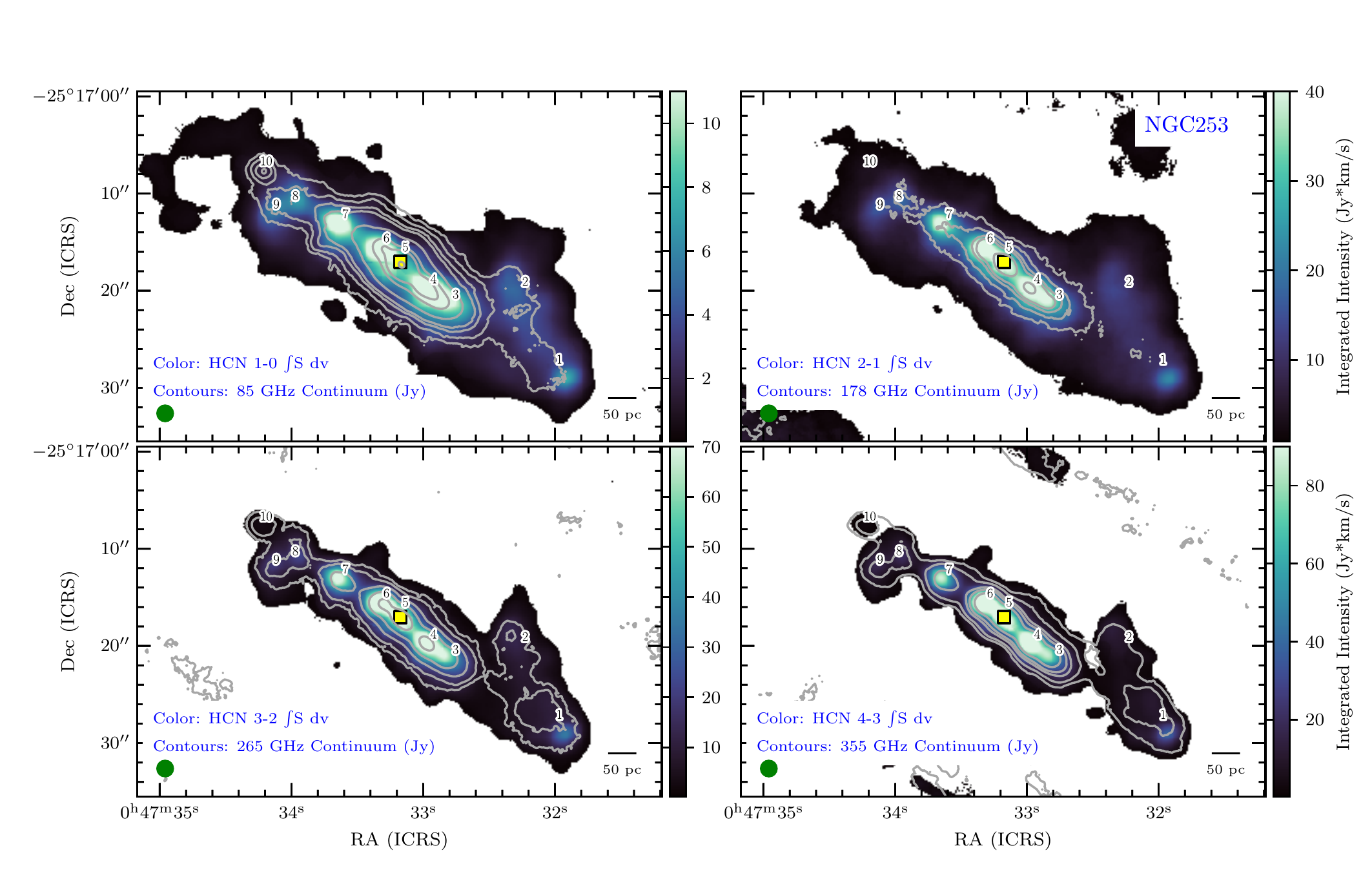}
\caption{HCN integrated intensity (moment 0) images toward NGC\,253. For each image the green circle in the lower-left corner shows the final imaged beam size (1.6\,arcsec). White-bordered numbers indicate the locations of the dense molecular emission regions identified by \citet[][Table 4]{leroy15}. A yellow black-bordered square locates the position of the strongest radio continuum emission peak identified by \citet[][TH2: R.A.(J2000) = $00^{h}47^{m}33^{s}.18$, Dec(J2000) = $-25^\circ17^\prime16.^{\prime\prime}93$]{Turner1985}}. A scale bar in the lower-right of each panel provides the physical scale in parsecs for each image.  The lower integrated intensity limit for each transition is set to 3$\sigma$ (see Table \ref{tab:transitions}). Overlain in contours is the associated continuum emission distribution for each transition.  Continuum contours are in steps of 3, 6, 9, 12, 30, 120, 240, and 900 times the respective continuum RMS, where the peak continuum intensity dictates the number of these levels actually used for a given panel. The respective continuum RMS values for the transitions shown are 0.07, 1.5, 0.3, and 1.0\,mJy/beam.
\label{fig:HCNMom0}
\end{figure*}

\section{Observational Data} \label{sec:data}
\subsection{ALCHEMI Data}
\label{sec:alchemidata}

In the following we provide a summary of the observation setup used to acquire the ALCHEMI survey data. Full details regarding the data acquisition, calibration, and imaging are provided in \cite{ALCHEMI-ACA}. The ALMA Cycle 5 Large Program ALCHEMI (project code 2017.1.00161.L) imaged the CMZ within NGC\,253 in the ALMA frequency Bands 3, 4, 6, and 7. This survey was subsequently extended to Band 5 during ALMA Cycle 6 (project code 2018.1.00162.S). The nominal phase center of the observations is $\alpha(ICRS)$ = 00$^h$47$^m$33$^s$.26, $\delta(ICRS)$ = $-25^\circ$17$^\prime$17$^{\prime\prime}.7$. A common rectangular area which was $50^{\prime\prime} \times 20^{\prime\prime}$ ($850\times340$\,pc) at a position angle of $65^\circ$ (East of North) represented the nuclear region (CMZ) imaged in NGC\,253. The final angular and spectral resolution of the image cubes generated from these measurements were $1.^{\prime\prime}6$ ($\sim27$\,pc) and 8-9 km\,s$^{-1}$, respectively \citep{Martin19}. The combination of the 12\,m Array and Atacama Compact Array (ACA) measurements used in this analysis resulted in a common maximum recoverable angular scale of $15^{\prime\prime}$ at all frequencies.  The rest-frequency coverage of ALCHEMI ranged from 84.2 to 373.2\,GHz.

\begin{figure*}[htbp]
\includegraphics[trim = 5mm 5mm 5mm 5mm, scale=0.74]{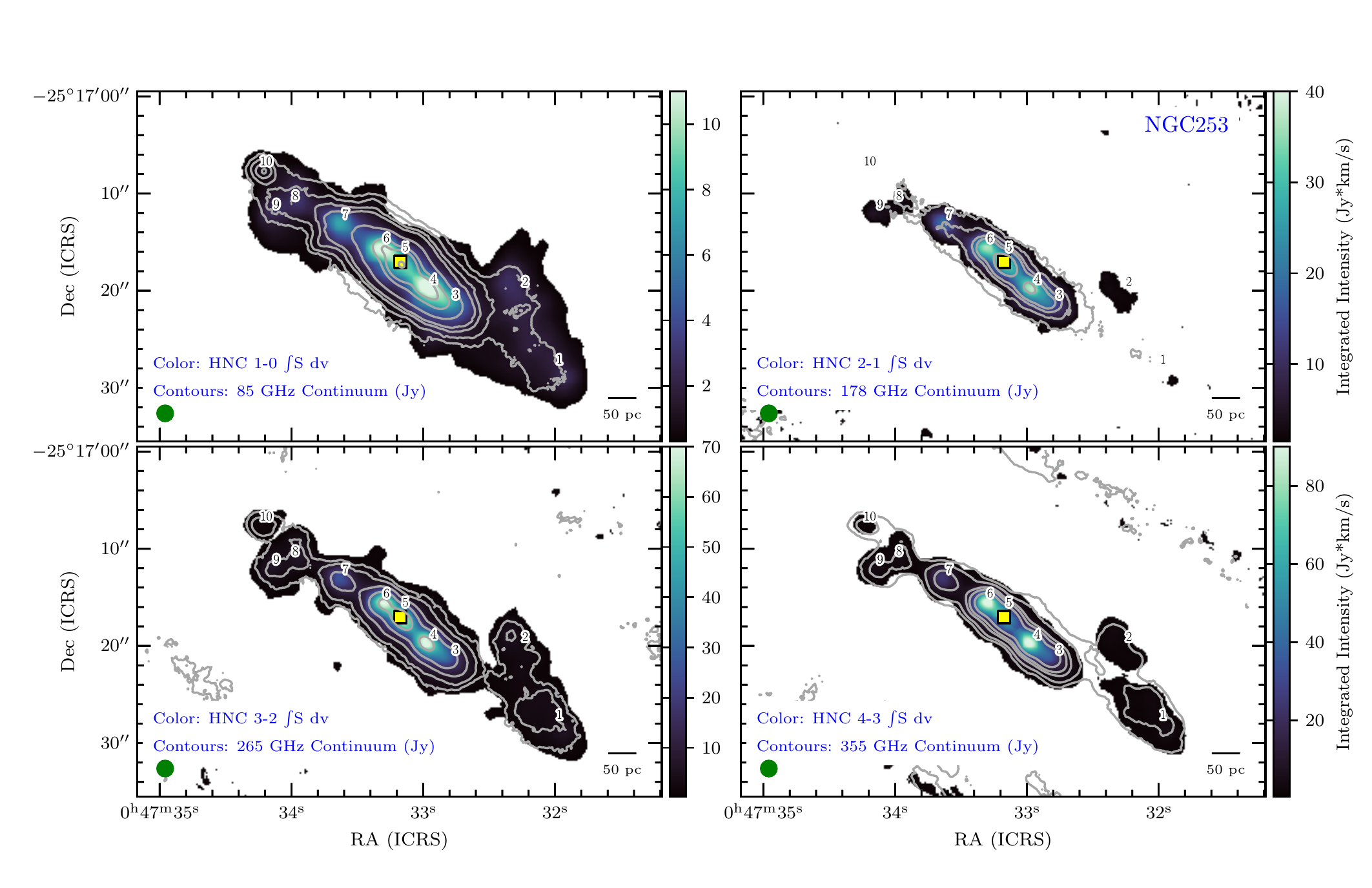}
\caption{HNC integrated intensity (moment 0) images toward NGC\,253.  Markings, intensity scaling, and contours in each panel are the same as for Figure~\ref{fig:HCNMom0}.}
\label{fig:HNCMom0}
\end{figure*}

%\subsection{ALMA %Data} %\label{sec:alma_re%s}

From the ALCHEMI archive we extract the $\sim 1.6$ arcsec resolution mosaics of the central molecular zone of NGC\,253 in the HCN and HNC $1-0$, $2-1$, $3-2$, and $4-3$ rotational transitions.  Table~\ref{tab:transitions} lists the transitions, frequencies, and spectral channel RMS values for all measurements studied.  We also extract the continuum emission associated with the measurements listed in Table~\ref{tab:transitions}.  The continuum subtraction and imaging processes used in this analysis are described in \cite{ALCHEMI-ACA}.

\begin{deluxetable}{lcc}
%\tablewidth{0pt}
\tabletypesize{\footnotesize}
\tablecolumns{3}
\tablecaption{HCN Isomer Measurements\label{tab:transitions}}
\tablehead{
\colhead{Transition} &
\multicolumn{2}{c}{HCN, HNC}
\\[-5pt]
\colhead{$J - (J-1)$} &
\colhead{Frequency} &
\colhead{$\sigma_{chan}$}
\\[-7pt]
\colhead{} &
\colhead{(GHz)} &
\colhead{(mJy/beam)} 
}
\startdata
$1-0$  &  88.632, 90.664 & 0.27, 0.26 \\
$2-1$ & 177.261, 181.325 & 6.41, 12.52 \\
$3-2$  & 265.886, 271.981 & 1.46, 1.98 \\
$4-3$  & 354.505, 362.630 & 2.70, 3.47 \\
\enddata
\end{deluxetable}

\subsection{Spectral Line Signal Extraction} \label{sec:extraction}

In order to extract integrated spectral line intensities from our measurements we use the \texttt{CubeLineMoment}\footnote{\url{https://github.com/keflavich/cube-line-extractor}} script introduced for this same purpose by \cite{mangum19}.  \texttt{CubeLineMoment} uses a series of spectral and spatial masks to extract integrated intensities for a defined list of target spectral frequencies.  As noted by \cite{mangum19}, the \texttt{CubeLineMoment} masking process uses a bright spectral line whose velocity structure is representative of the emission over the galaxy as a ``tracer" of the gas under study.  As the HCN and HNC emission measured toward NGC\,253 is quite intense in all transitions we were able to use each as its own tracer.  Final moment 0 (integrated intensity; Jy km\,s$^{-1}$), 1 (average velocity; km\,s$^{-1}$) and 2 (velocity dispersion; km\,s$^{-1}$) images are generated using a signal limit of three-times the spectral channel baseline RMS for the respective transition under study.

\begin{figure*}[htbp]
%\centering
%trim option's parameter order: left bottom right top
%\includegraphics[trim=0mm 10mm 0mm 0mm, clip, scale=0.305]{figures/NGC253-HCNHNC10Ratio.pdf}
%\includegraphics[trim=0mm 10mm 0mm 0mm, clip, scale=0.305]{figures/NGC253-HCNHNC21Ratio.pdf}\\
%\includegraphics[trim=0mm 0mm 0mm 0mm, clip, scale=0.305]{figures/NGC253-HCNHNC32Ratio.pdf}
%\includegraphics[trim=0mm 0mm 0mm 0mm, clip, scale=0.305]{figures/NGC253-HCNHNC43Ratio.pdf}
\includegraphics[trim = 5mm 5mm 5mm 5mm, scale=0.74]{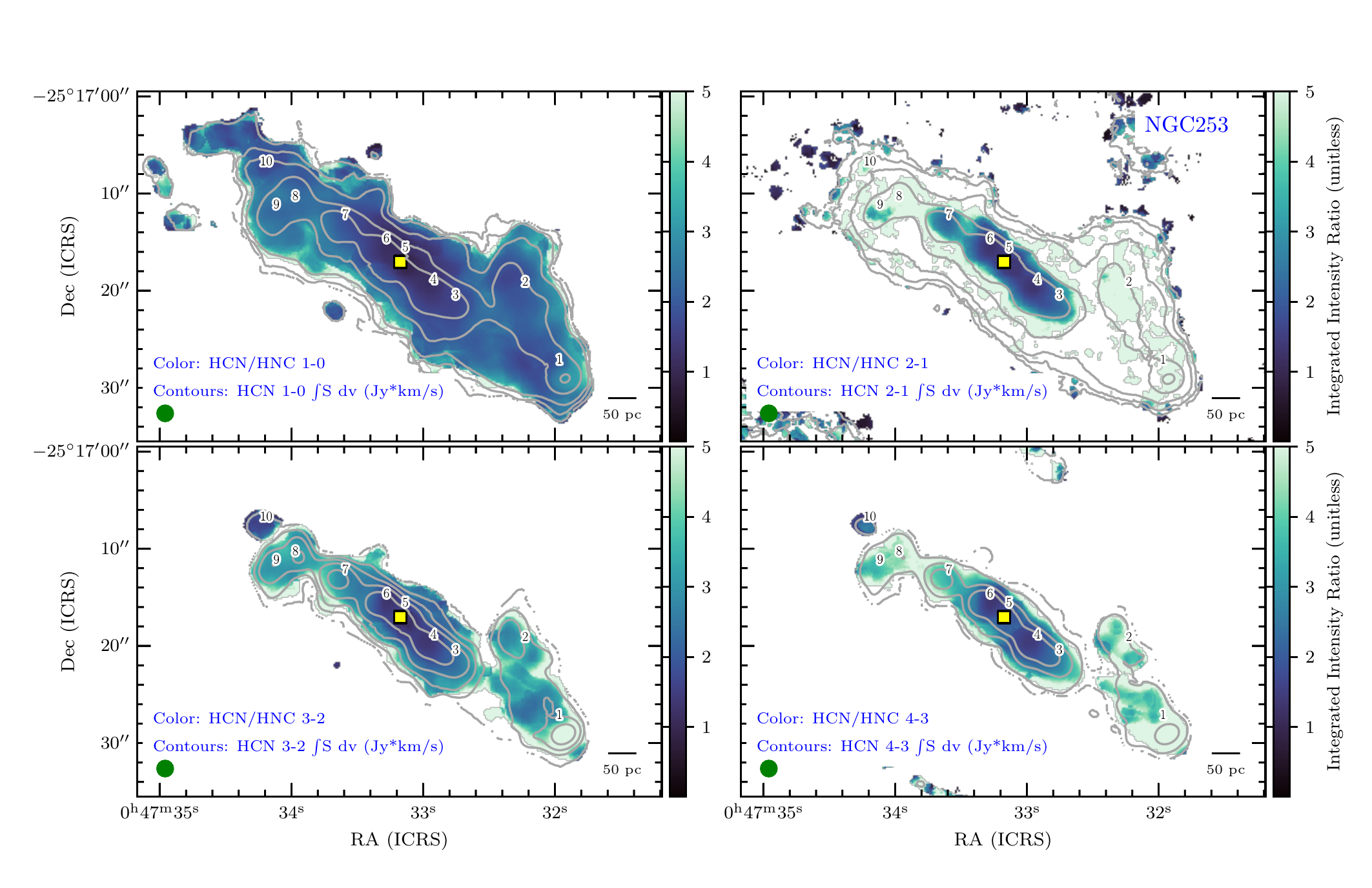}
\caption{HCN/HNC integrated intensity (moment 0) ratio images toward NGC\,253.  Contour levels are (0.2, 0.5, 1.0, 3.0, 7.0), (1.0, 2.0, 4.0, 10.0, 20.0), (1.0, 5.0, 10.0, 20.0, 50.0), and (1.0, 5.0, 20.0, 50.0) Jy/beam km\,s$^{-1}$ for the $1-0$, $2-1$, $3-2$, and $4-3$ HCN integrated intensities, respectively.  Markings in each panel same as for Figure~\ref{fig:HCNMom0}.}
\label{fig:HCNHNCRatio}
\end{figure*}

The moment 0 images for all HCN and HNC transitions are shown in Figures~\ref{fig:HCNMom0} and \ref{fig:HNCMom0}.  Ratios of each moment-0 HCN isomer for each transition have also been calculated (Figure~\ref{fig:HCNHNCRatio}).

To obtain integrated intensity values from across the CMZ while taking into account the limits of our resolution, we average the integrated intensity emission inside each of the 10 GMC-like structures identified by \cite{leroy15}. \cite{leroy15} define a GMC as an overdensity in molecular line emission on scales of $\sim$ 50\,pc. Using this definition, \cite{leroy15} identify 10 GMCs in the NGC\,253 CMZ (Table~\ref{tab:GMCpos}), though these clouds are noted to have higher densities ($n_{\text{H}_2} \sim 2000$\,cm$^{-3}$ over a three-dimensional GMC-sized FWHM)  and line widths ($\sigma \sim$ 20--40\,km\,s$^{-1}$) than GMCs found in our own Galaxy.  

We extract HCN and HNC integrated intensities for each of the four transitions and average each of them over these GMCs, adopting diameters equal to our beam size ($1.^{\prime\prime}6$, which is much smaller than the maximum recoverable angular scale of $15^{\prime\prime}$ for the ALCHEMI image cubes). Though we do not center these GMCs on peaks in the HCN and HNC emission, we find that the emission is smooth enough that any potential offset between the centers of the two species' emission would not substantially affect our calculated integrated intensities. Any remaining dilution of the HCN and HNC emission when averaging over the chosen GMC positions will underestimate the intensity of that emission. Uncertainties are calculated taking into account spectral channel RMS values, line widths, and absolute flux calibration uncertainties \citep{ALCHEMI-ACA} for each integrated intensity measurement. A list of the GMC-averaged integrated intensities is shown in Table~\ref{tab:obs_int}. 

These measurements suggest that the HCN/HNC integrated intensity ratio for all four transitions ranges from 1 to 5 (Figure~\ref{fig:HCNHNCRatio}), which is similar to that measured toward a wide range of galaxy types (Section~\ref{sec:gal_hcn}). These ratios are at their lowest ($\sim1-2$) in the central region of the CMZ, which encompasses GMCs 3--6.

\begin{deluxetable*}{ccccccccc}
\centering
\tablecolumns{9}
\tablewidth{0pt}
\tablecaption{GMC-Averaged Integrated Intensities\tablenotemark{a} \label{tab:obs_int}}
\tablehead{\colhead{GMC} & \colhead{HCN $1-0$} &\colhead{HCN $2-1$} & \colhead{HCN $3-2$} & \colhead{HCN $4-3$} & \colhead{HNC $1-0$} & \colhead{HNC $2-1$} & \colhead{HNC $3-2$} & \colhead{HNC $4-3$}}
\startdata
1 & 3.81(0.57) & 12.86(1.95) & 15.29(2.29) & 14.93(2.24) & 1.72(0.26) & 0.53(0.57) & 3.82(0.58) & 2.45(0.40) \\
2 & 4.23(0.63) & 13.43(2.03) & 11.30(1.70) & 8.56(1.23) & 2.22(0.33) & 2.04(0.64) & 3.35(0.51) & 1.71(0.30) \\
3 & 10.57(1.59) & 35.16(5.28) & 53.35(8.00) & 62.51(9.34) & 6.68(1.00) & 17.04(2.62) & 25.35(3.80) & 26.98(4.05) \\
4 & 13.27(1.99) & 	49.11(7.37) & 88.06(13.21) & 116.47(17.47) &	10.48(1.57)	& 34.04(5.14) &	63.52(9.53) &	77.60(11.64) \\
5 &	9.16(1.37) &	45.41(6.82) &	75.94(11.39) &	99.24(14.89)	& 8.05(1.21)	& 28.63(4.33)	& 53.41(8.01)	& 48.47(7.27) \\
6 &	12.68(1.90)	& 49.86(7.49) & 93.05(13.96) &	123.40(18.51)	& 9.84(1.48)	& 31.92(4.82) &	58.76(8.81) & 	73.21(10.98) \\
7 &	11.91(1.79) &	42.65(6.40) & 	67.05(10.06) &	74.06(11.11) &	5.90(0.89) &	16.98(2.61) &	22.78(3.42) &	19.91(2.99) \\
8 &	5.37(0.81)	& 15.52(2.35) &	16.36(2.45) &	13.00(1.95) &	2.32(0.35) &	2.77(0.70)	& 5.08(0.77) &	3.05(0.48) \\
9 &	4.74(0.71) &	14.53(2.20) &	16.34(2.45) &	14.10(2.12) &	2.05(0.31) &	4.01(0.82) &	5.41(0.82) &	3.90(0.61) \\
10 &	1.48(0.22) &	3.74(0.63) &	2.78(0.42) &	1.93(0.31) &	0.76(0.11) &	0.58(0.57)	& 1.65(0.26) &	0.85(0.20) \\
\enddata
\tablenotetext{a}{All integrated intensities have units of Jy\,km\,s$^{-1}$ with 1$\sigma$ uncertainties shown within parentheses.}
\end{deluxetable*}

%Figure~\ref{fig:IsomerRatio} shows the results of %these isomer ratio calculations.

%Two notable trends can be seen in %Figure~\ref{fig:IsomerRatio}:
%\begin{itemize}
%    \item For each GMC, the $^{12}$C$^{14}$N %isotopologue ratio increases as one goes from low %to high transition excitation.
%    \item For each GMC, the ratio increases as %one goes from $^{12}$C$^{14}$N to %$^{13}$C$^{14}$N to $^{12}$C$^{15}$N isotopologue %ratios.
%\end{itemize}
%In Section~\ref{sec:opticaldepth} we investigate %possible interpretations of these integrated %intensity ratio trends within our radiative %transfer modeling.

%\begin{figure*}
%    \centering
%    \includegraphics[scale=0.6]{IsomerRatioPlotAl%l3sig-region_average-20220106.pdf}
%    \caption{Ratio of HCN to HNC isotopologue %spectral line integrated intensities $I$ toward %the 10 NGC\,253 GMCs.  A cutoff of $3\sigma$ in %integrated intensity has been applied to the %integrated intensity ratios shown.}
%    \label{fig:IsomerRatio}
%\end{figure*}

\subsection{Interloper Analysis} \label{sec:interloper}

\begin{figure}
%\centering
%trim option's parameter order: left bottom right top
%\gridline{
%\fig{figures/HC15N10Interloper.pdf}{0.45\textwidt%h}{(a)}
%\fig{figures/HC15N21Interloper.pdf}{0.45\textwidt%h}{(b)}
%}
%\gridline{
%\fig{figures/HC15N32Interloper.pdf}{0.45\textwidt%h}{(c)}
%\fig{figures/HC15N43Interloper.pdf}{0.45\textwidt%h}{(d)}
%}
%\gridline{
%\fig{figures/HNC43Interloper.pdf}{0.5\textwidth}{%(e)}
%}
\includegraphics[scale=0.5]{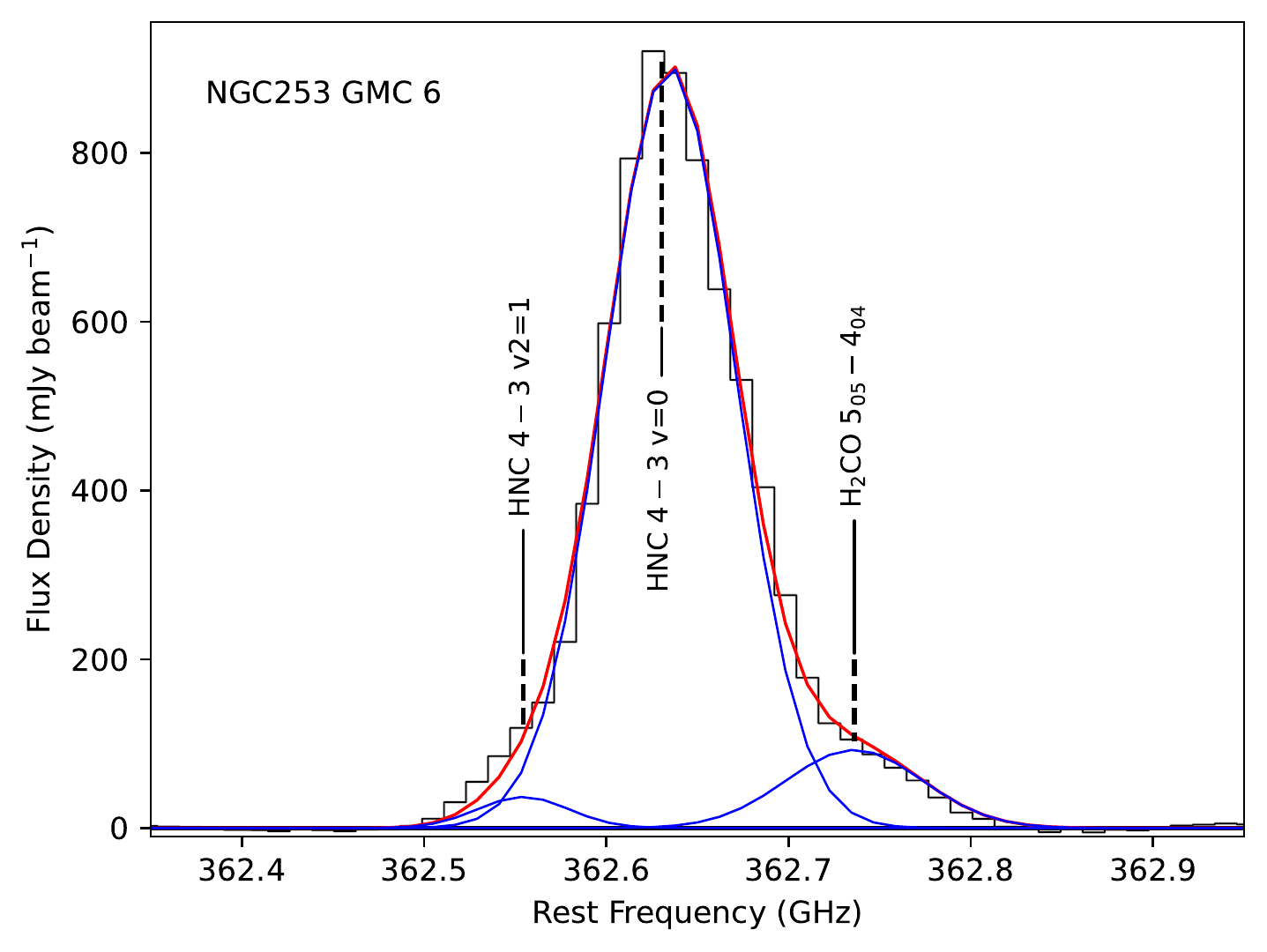}
\caption{Spectral interloper diagnostic spectrum associated with HNC $4-3$.  Individual (blue lines) and cumulative (red lines) Gaussian fits are shown.}
\label{fig:interloper}
\end{figure}

Our \texttt{CubeLineMoment} analysis includes a sample spectrum check to reveal potential spectral line blending. Only the HNC $4-3$ transition is found to have spectral neighbors which required assessment of the amount of emission contributed by H$_2$CO $5_{05}-4_{04}$ (362.736048\,GHz) and HNC $4-3$ v$_2$=1 (362.554351\,GHz). Using the procedure described in \cite{Holdship2022} we determine that these two interlopers contribute respectively at most 4\% and 1\% to the HNC $4-3$ integrated emission. This contamination estimate is consistent with the multi-species LTE analysis of molecular column densities described in \cite{ALCHEMI-ACA}. Figure~\ref{fig:interloper} shows a sample spectrum toward a central region in the NGC\,253 CMZ (Region 6) which indicates the spectral line blending of the HNC $4-3$ transition. Since the estimated correction required to this single transition is small, we do not apply these corrections to our presented integrated intensities.

%\begin{deluxetable*}{llll}[hb]
%\tablewidth{0pt}
%\tablecolumns{4}
%\tablecaption{HCN Isotopomer Interloper %Analysis\label{tab:interloper}}
%\tablehead{
%\colhead{Isotopomer and Transition} & 
%\colhead{Interloper and Frequency} & 
%\colhead{Distance} &
%\colhead{Impact} \\[-5pt]
%& \colhead{(GHz)} & \colhead{(MHz/\kms)} &
%}
%\startdata
%HC$^{15}$N $1-0$ & SO $2(2)-1(1)$ (86.093950) & %40/140 & $\sim 5$\% \\
%HC$^{15}$N $2-1$ & SO $4(4)-3(3)$ (172.181460) & %73/127 & $\sim 2$\% \\
%HC$^{15}$N $3-2$ & SO $6(6)-5(5)$ (258.255813) & %100/110 & $\sim 5$\% \\
%HC$^{15}$N $4-3$ & SO $8(8)-7(7)$ (344.310612) & %100/100 & $\sim 12$\% \\
%HNC $4-3$ v=0 & H$_2$CO $5_{05}-4_{04}$ %(362.736048) & 106/106 & $\sim 4$\% \\
%& HNC $4-3$ v$_2$=1 (362.554351) & 76/76 & $\sim %1$\% \\
%\enddata
%\end{deluxetable*}

\section{Coupled Radiative Transfer -- Chemical Modeling} %\label{sec:results}
\label{sec:model_methods}

\subsection{Model Description} \label{model_description}
We model the chemical and physical conditions within each of the 10 GMCs using the chemical modeling code \texttt{UCLCHEM}\textbf{\footnote{\url{https://uclchem.github.io/}}} \citep{uclchem} and the radiative transfer code \texttt{SpectralRadex}\footnote{\url{https://spectralradex.readthedocs.io}}.  

\texttt{UCLCHEM} is a gas-grain chemical modeling code that incorporates user-defined chemical networks to produce chemical abundances given the input physical conditions of the gas (e.g. gas temperature, volume density). We take our gas phase network from UMIST12 \citep{McElroy2013}, which includes two-body reactions between species as well as reactions with UV photons and cosmic rays, and use depleted abundances from Table 4 of \cite{jenkins2009} for our initial conditions. The cosmic ray reaction rates use a cosmic ray ionization rate of $\zeta_0 = 1.36\times10^{-17}$s$^{-1}$ from which the cosmic ray ionization rates are scaled. We augment this database by including the reaction,
\begin{equation}
    \text{HNC} + \text{O} \longrightarrow \text{CO} + \text{NH},
\end{equation}
which has been shown to be important in the chemistry of HCN and HNC \citep{hacar20}. The isomerization reaction that converts HNC into HCN is already included in the database:

\begin{equation}
    \text{HNC} + \text{H} \longrightarrow \text{HCN} + \text{H}.
\end{equation}
We test both the high (2000\,K, 1200\,K) and low (20\,K, 200\,K) barrier values \citep{hacar20} for the HNC + O and HNC + H reactions in order to understand the effects of temperature barriers on our modeling results. We further include \texttt{UCLCHEM}'s default grain surface reactions including freeze out, non-thermal desorption, and diffusive reactions between species adsorbed to the grain. We use a single point model to replicate the environment in the GMCs by assuming these gas clouds are homogeneous because they have high enough visual extinctions such that they are shielded from UV radiation \citep{harada21}. We calculate the species column density using the on-the-spot approximation, where we multiply the fractional abundance at the source of the emission by our H$_{2}$ column density \citep{Dyson1997}.

To incorporate radiative transfer modeling, we use \texttt{SpectralRadex}, a python library which includes a wrapper for the \texttt{RADEX}\footnote{\url{https://home.strw.leidenuniv.nl/~moldata/radex.html}} \citep{radex} program. \texttt{RADEX} is a 1D non-Local Thermodynamic Equilibrium (LTE) statistical equilibrium radiative transfer code that assumes an isothermal and homogeneous environment. Optical depth effects are treated within \texttt{RADEX} using an escape probability method. \texttt{RADEX} allows the user to do radiative transfer calculations while constraining physical conditions such as density and temperature. Given \texttt{UCLCHEM} chemical abundances and user-defined temperature, density, and H$_{2}$ column density values, we can use \texttt{RADEX} to connect chemical abundances to integrated intensities through the molecular column densities. We can then directly compare the model-predicted integrated intensities to our measurements. These integrated intensities are calculated assuming a uniform line width of 100 km\,s$^{-1}$, which is consistent with the line widths derived from our spectral line extraction procedure (Section \ref{sec:extraction}) and a beam-filling factor of 1. It is also important to note that we only consider excitation through collisions with H$_{2}$ and therefore ignore electron collisions. A previous ALCHEMI-based study \citep{Holdship2022} found that even at the cosmic ray ionization rates which will be discussed later in this article (Section \ref{sec:mod_res}), almost all hydrogen is in its molecular form under these conditions. \cite{Holdship2022} found that toward GMCs 3 through 7 that the fractional abundance of electrons is in the range X(e$^{-1}$) $\sim 10^{-4}-10^{-5}$ for volume densities n(H$_2$) $\lesssim 10^{5.5}$\,cm$^{-3}$.  \cite{Goldsmith2017ApJ} note that electrons could be of practical importance for HCN excitation when n(H$_2$) $< 10^{5.5}$\,cm$^{-3}$ and X(e$^{-1}$) $> 10^{-5}$. Even though electron-induced collisions could be important in the lower-density regions within the NGC\,253 CMZ, we have opted to not consider electron-induced collisions in our analysis, and to defer further analysis of the potential impact of electron collisions in our model to a future analysis.

As noted by \cite{Aalto2007A&A} the ground state vibrational energy levels of the HCN and HNC isomers can be populated via infrared excitation of the lowest-energy vibrational energy levels. This mechanism involves absorption of infrared photons by coupling to the lowest-energy (v$_2$=1) degenerate vibrational bending mode of each isomer. As described by \cite{Aalto2007A&A}, this infrared coupling has the effect of exciting the ground vibrational states of the HCN and HNC isomers to higher rotational levels via a $\Delta$J=2 selection rule. The v$_2$=1 bending modes in HCN and HNC have wavelengths of 14 and 22\,$\mu$m (714 and 464\,cm$^{-1}$, respectively), while their energies above ground (E$_{IR}$) are 1027 and 669\,K, respectively. The Einstein-A coefficients for these vibrational bending modes are A$_{IR}$ = 1.7 and 5.2\,s$^{-1}$ for HCN and HNC, respectively. Since the rate of an infrared pumped vibrational transition is given by $P_{IR} \propto A_{IR}/\exp(E_{IR}/T_{IR})$, where T$_{IR}$ is the infrared brightness temperature, the HNC infrared pump is approximately two orders of magnitude faster than that for HCN. This difference in infrared pumping efficiency results in an HCN/HNC ground vibrational state spectral line intensity ratio that is less than one. Since we do not measure spectral line intensity ratios less than 1 toward the NGC\,253 CMZ (Section~\ref{sec:extraction}), we did not see a justification for including infrared excitation in our radiative transfer model. This does not mean that infrared excitation of the ground vibrational energy states of HCN and HNC do not exist in the NGC\,253 CMZ, but that it is not a necessary excitation mechanism to explain our observations.

\begin{deluxetable*}{cccc}
\centering
\tablecolumns{3}
\tablewidth{0pt}
\tablecaption{Prior Distributions\label{tab:priors}}
\tablehead{
\colhead{} &  \colhead{Parameter} & \colhead{Range} & Distribution Type 
}
\startdata
$T$  & Temperature & 50--300 K & Uniform \\ [5pt]
$n$  & Volume Density & 10$^3$--10$^7$ cm$^{-3}$ & Log-uniform \\ [5pt]
$\zeta$ & Cosmic Ray Ionization Rate & 10--10$^7$ $\zeta_{0}$\tablenotemark{a} & Log-uniform \\[5pt]
$N_{\text{H}_2}$ & H$_{2}$ Column Density & 10$^{22}$--$10^{25}$ cm$^{-2}$ & Log-uniform\\[5pt]
\enddata
\tablenotetext{a}{$\zeta_0$ = $1.36\times10^{-17}$\,s$^{-1}$}
\end{deluxetable*}

\begin{deluxetable*}{ccccccccccc}
\centering
\tablecolumns{9}
\tablewidth{0pt}
\tablecaption{NGC\,253 GMC Physical Parameters\tablenotemark{a}\label{tab:mod_results}}
\tablehead{
\colhead{GMC} & \colhead{} & \colhead{T$_\text{K}$} & \colhead{}& \colhead{$\log_{10}$ n} & \colhead{}& \colhead{$\log_{10}\zeta$} & \colhead{} &   \colhead{$\log_{10}$ N$_{\text{H}_2}$} \\[-7pt]
\colhead{} & \colhead{} & \colhead{[K]} & \colhead{} & \colhead{[cm$^{-3}$]} & \colhead{} & \colhead{[$\zeta_{0}$]} &  \colhead{} & \colhead{[cm$^{-2}$]}
} 
\startdata
1   & & 172.53$^{+75.77}_{-67.79}$ & & $3.81^{+0.90}_{-0.48}$  &  &  $3.87^{+0.15}_{-0.22}$ & & $22.85^{+0.82}_{-0.54}$  \\[5pt]
2    &   &  135.94$^{+69.52}_{-52.60}$      & & $3.89^{+0.85}_{-0.53}$  & &  $3.80^{+0.20}_{-0.07}$  &  & $<$ 23.79\\[5pt]
3   &  & 161.72$^{+93.81}_{-65.15}$ & & 4.73$^{+0.73}_{-0.88}$ & &    4.08$^{+ 0.81}_{-0.24}$ & &  23.24$^{+0.80}_{-0.79}$\\[5pt]
4  &   & NC & & 5.31$^{+0.48}_{-1.00}$ & &  4.82$^{+0.79}_{-0.87}$ & & NC \\[5pt]
5  &    &    NC &  & 5.62$^{+0.31}_{-0.34}$ &  & 5.09$^{+0.39}_{-0.50}$ & & 23.50$^{+0.87}_{-0.67}$\\[5pt]
6    &  &   NC   & & 5.43$^{+0.39}_{-0.81}$ & & 4.85$^{+0.61}_{-0.86}$ & & 23.41$^{+1.07}_{-0.86}$ \\[5pt]
7      &    &   148.13$^{+98.79}_{-60.97}$  & & 4.79$^{+0.68}_{-0.97}$ & &  3.97$^{+0.41}_{-0.16}$ & & $<$ 23.90 \\[5pt]
8   & & 162.11$^{+80.99}_{-68.79}$ &  & $3.92^{+0.68}_{-0.53}$ & &  3.90$^{+0.14}_{-0.28}$ & & $<$ 23.86 \\[5pt]
9  &  &  163.92$^{+77.62}_{-62.81}$   &    & $3.98^{+0.89}_{-0.58}$ & & 3.90$^{+0.16}_{-0.16}$ & & $<$ 23.79 \\[5pt]
10   &  &  NC   &    & $3.93^{+0.77}_{-0.58}$ &  & 4.15$^{+0.41}_{-0.46}$ & & $<$ 23.93\\
\enddata
\tablenotetext{a}{Most likely parameters describing each GMC as a result of \texttt{UCLCHEM} + \texttt{RADEX} modeling and \texttt{UltraNest} sampling. Uncertainties indicate $\pm$ 33\% of posterior distribution. $<$ indicates upper limit (83rd percentile) of distribution. NC = not constrained.}
\end{deluxetable*}

\begin{figure*}
    \centering
    \includegraphics[scale=0.55]{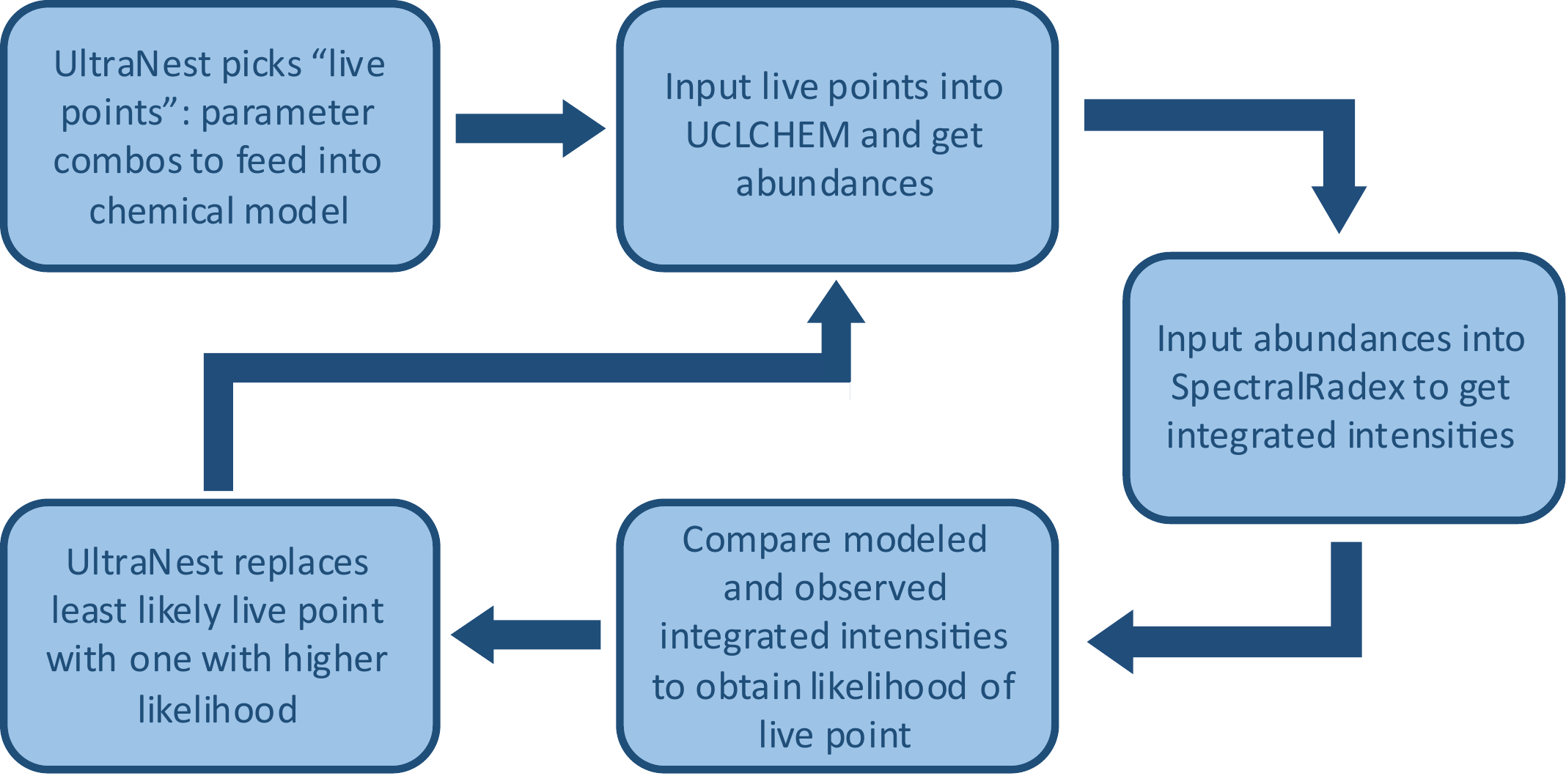}
    \caption{Flow chart describing our nested sampling + chemical and radiative transfer modeling process.}
    \label{fig:flow}
\end{figure*}

\subsection{Defining Bayesian Priors} \label{priors}
We are interested in estimating density, temperature, cosmic ray ionization rate, and molecular hydrogen column density in the NGC\,253 CMZ. Our choices for parameter prior distributions are listed in Table \ref{tab:priors} for volume density $n$, kinetic temperature $T$, cosmic ray ionization rate $\zeta$, and molecular hydrogen column density $N_{\text{H}_2}$. For our temperature parameter, we sample kinetic temperatures between 50 and 300\,K, adopting a flat prior distribution to uniformly sample the parameter space without bias. This kinetic temperature prior is based on the results of the \cite{mangum19} kinetic temperature measurements toward the NGC\,253 CMZ.  On the largest angular scales ($\sim5^{\prime\prime}$), \cite{mangum19} measured kinetic temperatures $\sim 50$\,K. On smaller scales ($\lesssim1^{\prime\prime}$), \cite{mangum19} measured kinetic temperatures of at least 300\,K.

We model cosmic ray ionization rates with a log-uniform distribution ranging from 10\,$\zeta_{\text{0}}$ -- $10^{7}\,\zeta_{\text{0}}$ ($\sim 10^{-16} - 10^{-10}$ s$^{-1}$). We adopt this upper limit by taking into consideration estimates made by \cite{holdship_c2h} and \cite{harada21}, which derive $\zeta$ ranges from $10^{3}-10^{6}\,\zeta_{\text{0}}$. Since \cite{harada21} estimates one general CRIR in the CMZ and \cite{holdship_c2h} only analyzes GMCs 3--7, we have no point of reference for outer GMCs 1, 2, 8, 9, and 10. Thus, we adopt a lower limit of 10\,$\zeta_{0}$ to account for a potentially low CRIR in these less active regions.

We adopt a log-uniform distribution for densities over the range 10$^{3} - 10^{7}$ cm$^{-3}$. Observations suggest gas densities of $10^{5} - 10^{6}$ cm$^{-3}$ \citep{leroy18,harada21}, so we model densities centered on this range with a few orders of magnitude as a buffer both higher and lower than this estimate.

For our molecular hydrogen column density prior, we rely upon previous measurements of this quantity toward the CMZ of NGC\,253.  Millimeter dust continuum measurements over similar spatial scales as those modelled here were used to derive N$_{\text{H}_2}$ in the range $10^{23}$ to $7\times10^{24}$\,cm$^{-2}$ for GMCs\,3 through 7 \citep{mangum19}. From these measurements we set the upper-bound of our N$_{\text{H}_2}$ prior to $10^{25}$\,cm$^{-2}$. Since the \cite{mangum19} measurements did not sample GMCs\,1, 2, 8, 9, or 10, which appear to be in regions of lower dust column density (Figure~\ref{fig:ua_srcs}), we have adopted $10^{22}$\,cm$^{-2}$ for the lower-bound of our N(H$_2$) prior. Again, we use a log-uniform distribution for this prior.

\begin{figure*}
    \centering
    \includegraphics[trim= 20mm 0mm 20mm 0mm, scale=0.8]{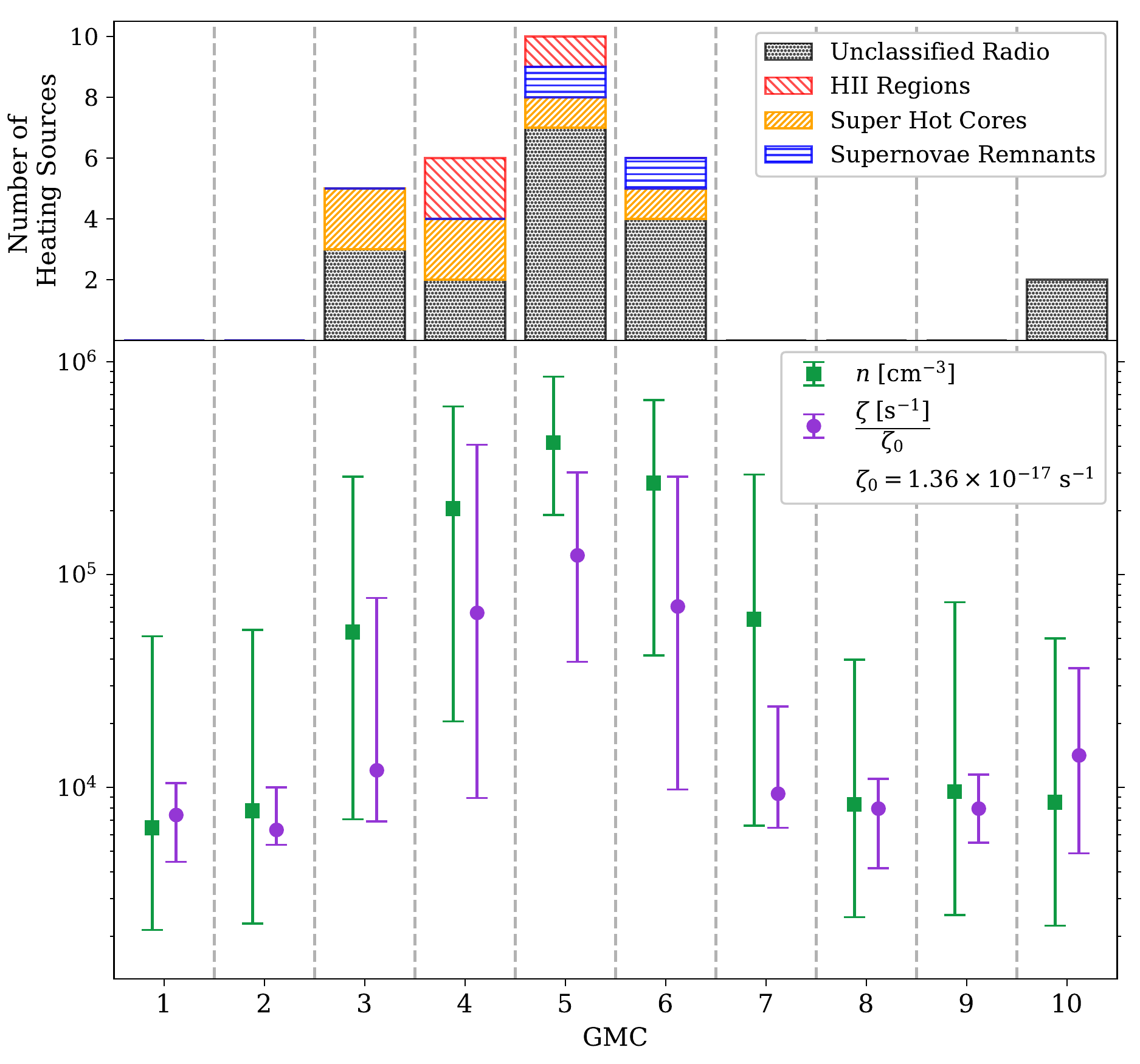}
    \caption{Top: Number of heating sources per GMC. ``Unclassified", ``Supernovae Remnants", and ``HII Regions" are from \cite{ua97}, while ``Super Hot Cores" are from \cite{RV2020}. Note that the Super Hot Core source measurements sample only the inner portion of the NGC\,253 CMZ, which includes GMCs 3 through 6. Bottom: Median modeled volume density (green squares) and cosmic ray ionization rate (purple circles) values for each GMC. Error bars indicate the 16th--84th percentile of the posterior distributions.}
    \label{fig:samp_results}
\end{figure*}

\begin{figure}
    \centering
    \includegraphics[trim=5mm 8mm 0mm 0mm, scale=0.7]{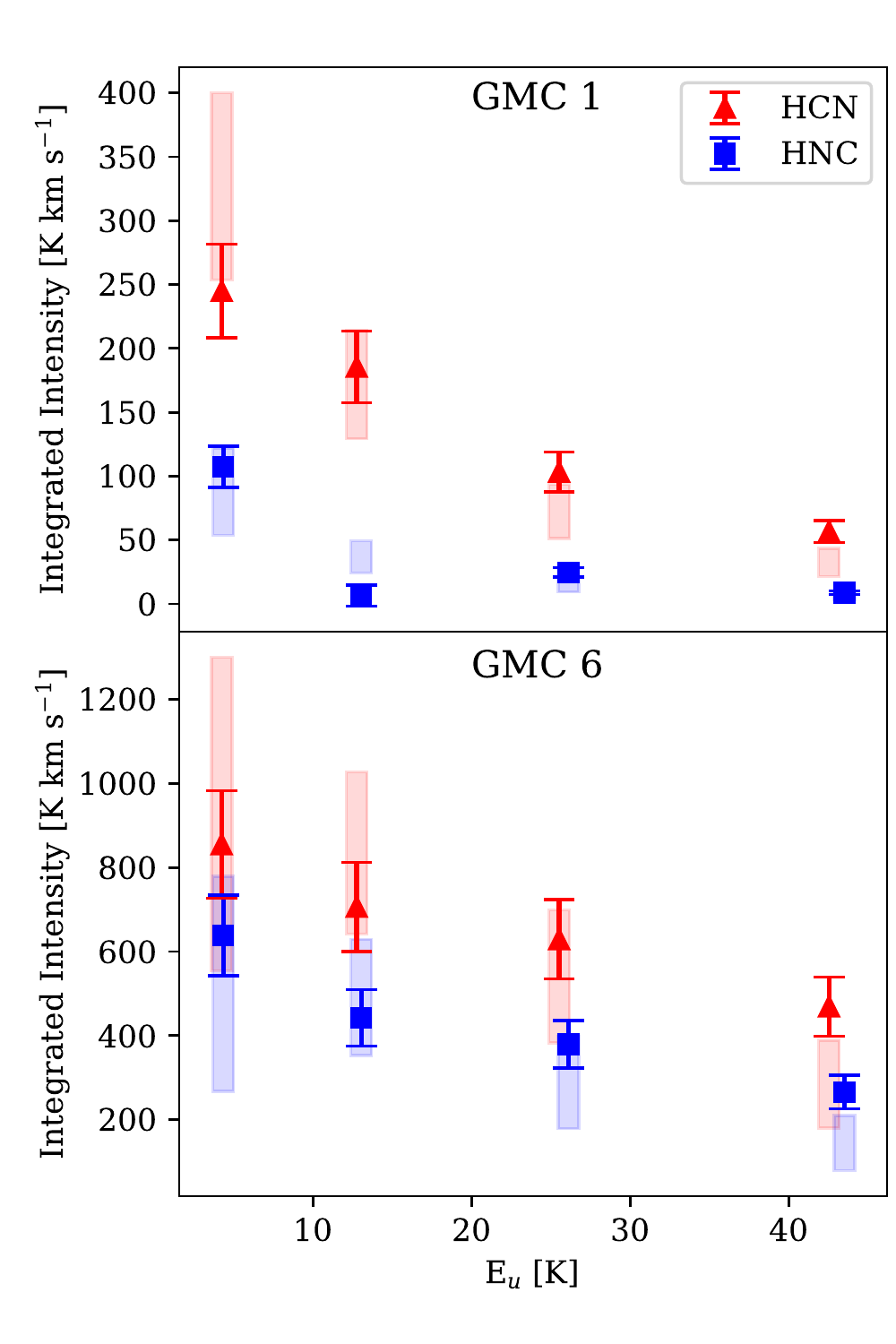}
    \caption{Observed (triangles and squares) versus modeled (shaded bars) flux for HCN (red) and HNC (blue). Observed error bars indicate the 1$\sigma$ uncertainty range. Shaded rectangles show the inner 67\% ($\sim$16th---84th percentile) of our modeled flux distributions.}
    \label{fig:fluxes}
\end{figure}

\begin{figure*}
    \centering
    \includegraphics[trim= 3mm 0mm 0mm 0mm, scale=0.63]{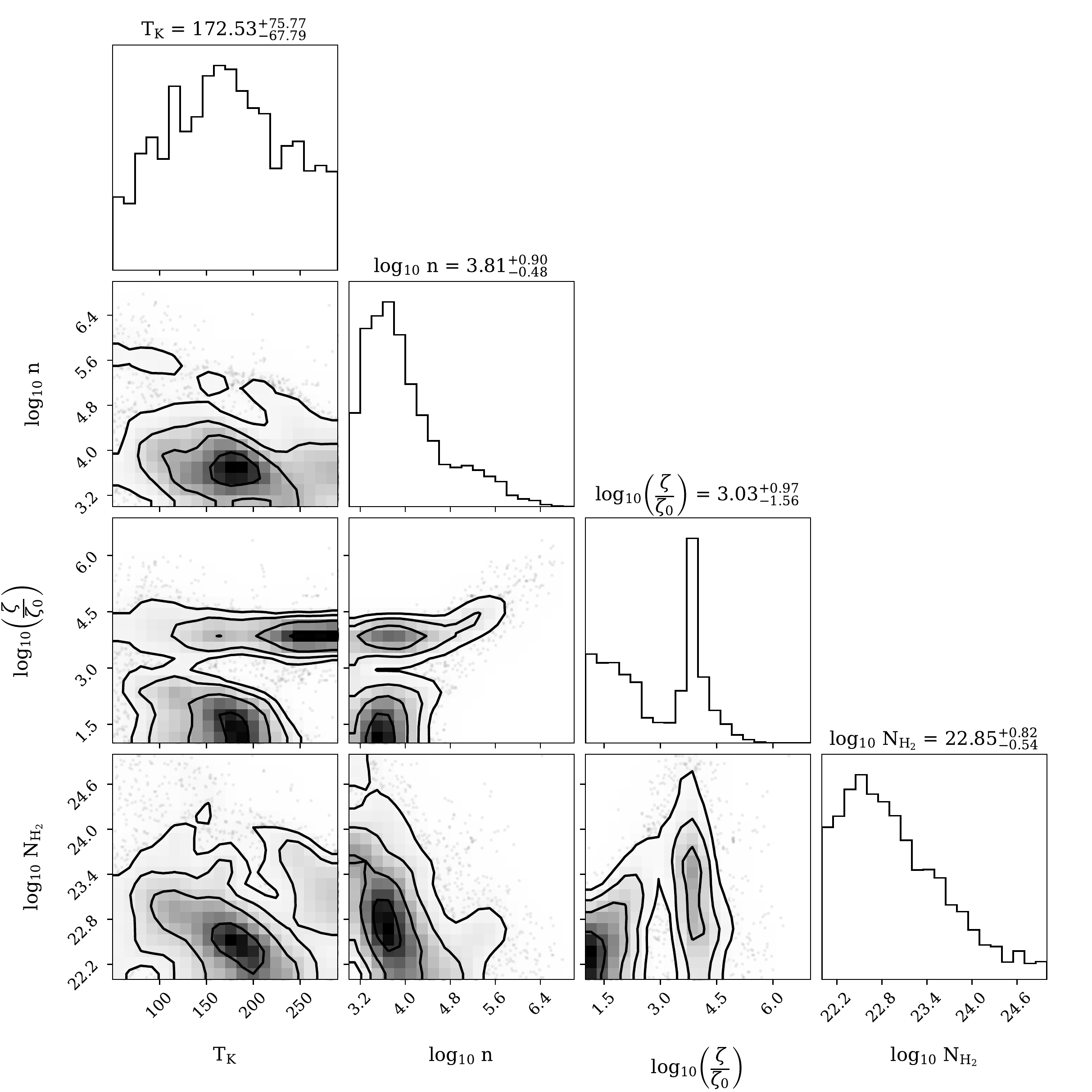}
    \caption{Modeling results for GMC 1.}
    \label{fig:corner1}
\end{figure*}

\begin{figure*}
    \centering
    \includegraphics[trim = 3mm 0mm 0mm 0mm, scale=0.63]{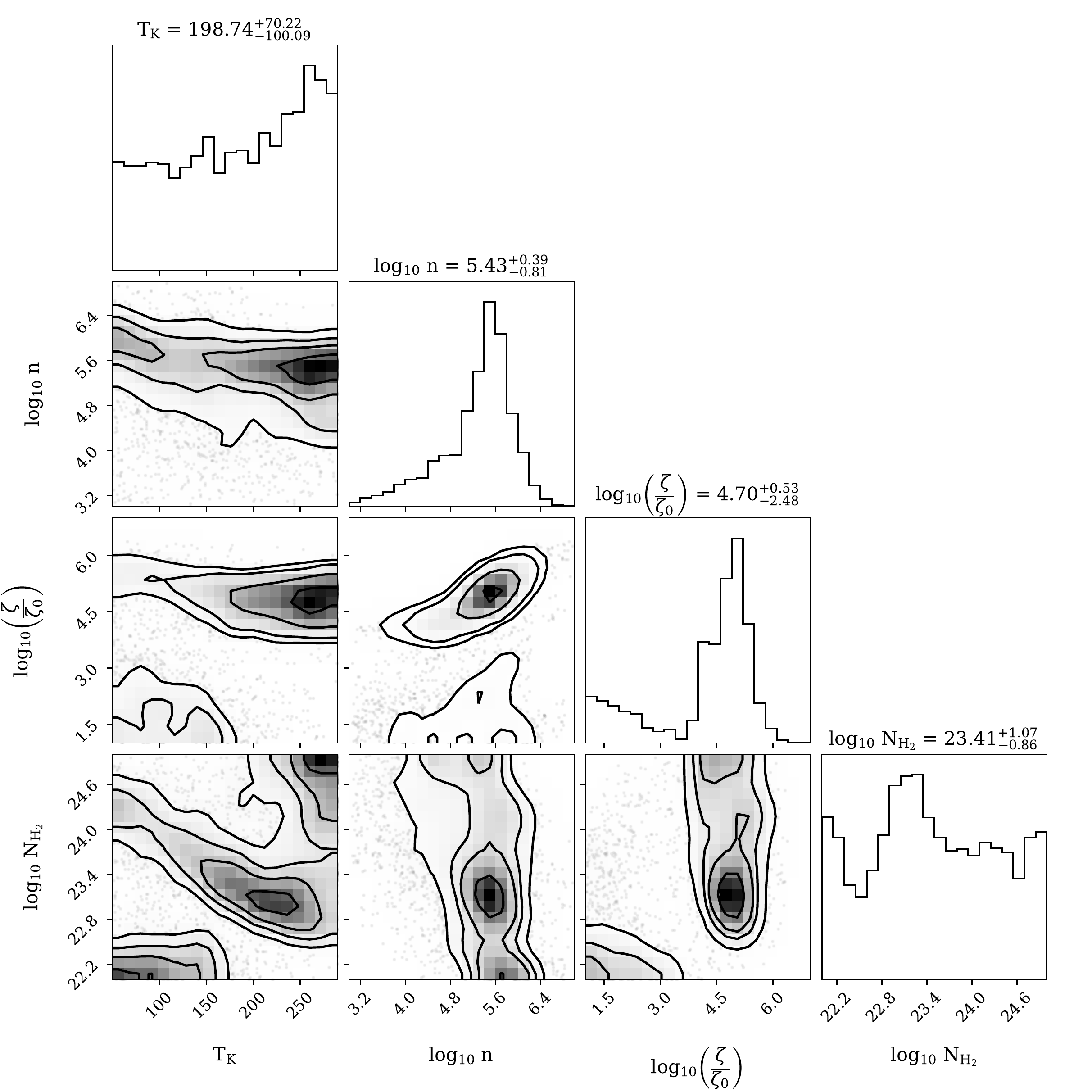}
    \caption{Modeling results for GMC 6.}
    \label{fig:corner6}
\end{figure*}

\subsection{Nested Sampling} \label{nest}

We sample our parameter space to obtain input for our chemical models using nested sampling techniques by implementing the Monte Carlo algorithm MLFriends \citep{ultranest14,ultranest19} using the \texttt{UltraNest}\footnote{\url{https://johannesbuchner.github.io/UltraNest/}} package \citep{ultranest21}. \texttt{UltraNest}'s MLFriends algorithm estimates the posterior probability distribution of some parameters given our data, using Bayes' theorem

\begin{equation}
    P(\boldsymbol\theta | \mathbf{F_d}) = \frac{P(\mathbf{F_d}|\boldsymbol\theta) P(\boldsymbol\theta)}{P(\mathbf{F_d})},
        \label{eq:param_like}
\end{equation}
where $P(\mathbf{F_{d}}|\boldsymbol\theta)$ is the probability of obtaining some data $F_{d}$ given a set of parameters $\theta$ (e.g. $T$, $n$), $P(\boldsymbol\theta)$ is the prior probability of those parameters, and $P(\mathbf{F_{d}})$ is the Bayesian evidence.

We can determine $P(\mathbf{F_{d}}|\boldsymbol\theta)$ by assuming Gaussian errors giving the standard function
\begin{equation}
    P(\mathbf{F_d}|\boldsymbol{\theta}) = \exp\left(-\frac{\displaystyle 1}{\displaystyle 2} \sum\limits_{i} \frac{(F_{d,i} - F_{t,i})^2}{\sigma^2_{F,i}}\right)
    \label{eq:lik_fn}.
\end{equation}
In Equation \ref{eq:lik_fn}, we compare our data $F_d$ and its uncertainty $\sigma_{F}$ to the output of our forward model $F_t$ for any given set of parameters that we obtain for each transition $i$. 

To sample the posterior distribution, \texttt{UltraNest} initially samples the entire parameter space by selecting a number of parameter combinations, called ``live points", based on the prior probability of our parameters, and then replacing the least likely of these combinations based on the results from chemical and radiative transfer modeling. As shown graphically in Figure \ref{fig:flow}, in each iteration, the selected parameters are fed into \texttt{UCLCHEM}, producing chemical abundances of the desired HCN and HNC transitions as a fraction of total H nuclei. We combine modeled abundances with molecular hydrogen column density as a free parameter to obtain HCN and HNC column densities. We input these values into \texttt{SpectralRadex} to obtain integrated intensities to compare to our ALMA observations of the HCN and HNC 1--0, 2--1, 3--2, and 4--3 transitions. \texttt{SpectralRadex} produces integrated intensities in K\,km\,s$^{-1}$, so we can use the following equation to convert our observed beam-averaged integrated intensities from Jy\,km\,s$^{-1}$ to K\,km\,s$^{-1}$:
\begin{multline}
    T_R(\text{K}) = 13.59\left(\dfrac{300 \text{GHz}}{\nu}\right)^2 \\ \times \left(\dfrac{1^{\prime\prime}}{\theta_{max}}\right)\left(\dfrac{1^{\prime\prime}}{\theta_{min}}\right)I(\text{Jy}),
\end{multline}
where $\nu$ is the rest frequency of the line, $\theta_{max}$ and $\theta_{min}$ are the FWHMs of the major and minor axes of our Gaussian beam, and $I$ is our integrated intensity. In our case $\theta_{max} = \theta_{min} = 1.^{\prime\prime}6$. At each iteration, the live point with the lowest likelihood is removed and replaced with a more suitable point, which results in the volume of the sampled parameter space shrinking. These iterations continue until the live point weights are insignificant (fractional remainder $\leq$ 0.01), indicating the vast majority of the probability density has been sampled.

\begin{figure*}
    \centering
    \includegraphics[trim = 2mm 5mm 0mm 5mm, scale=0.7]{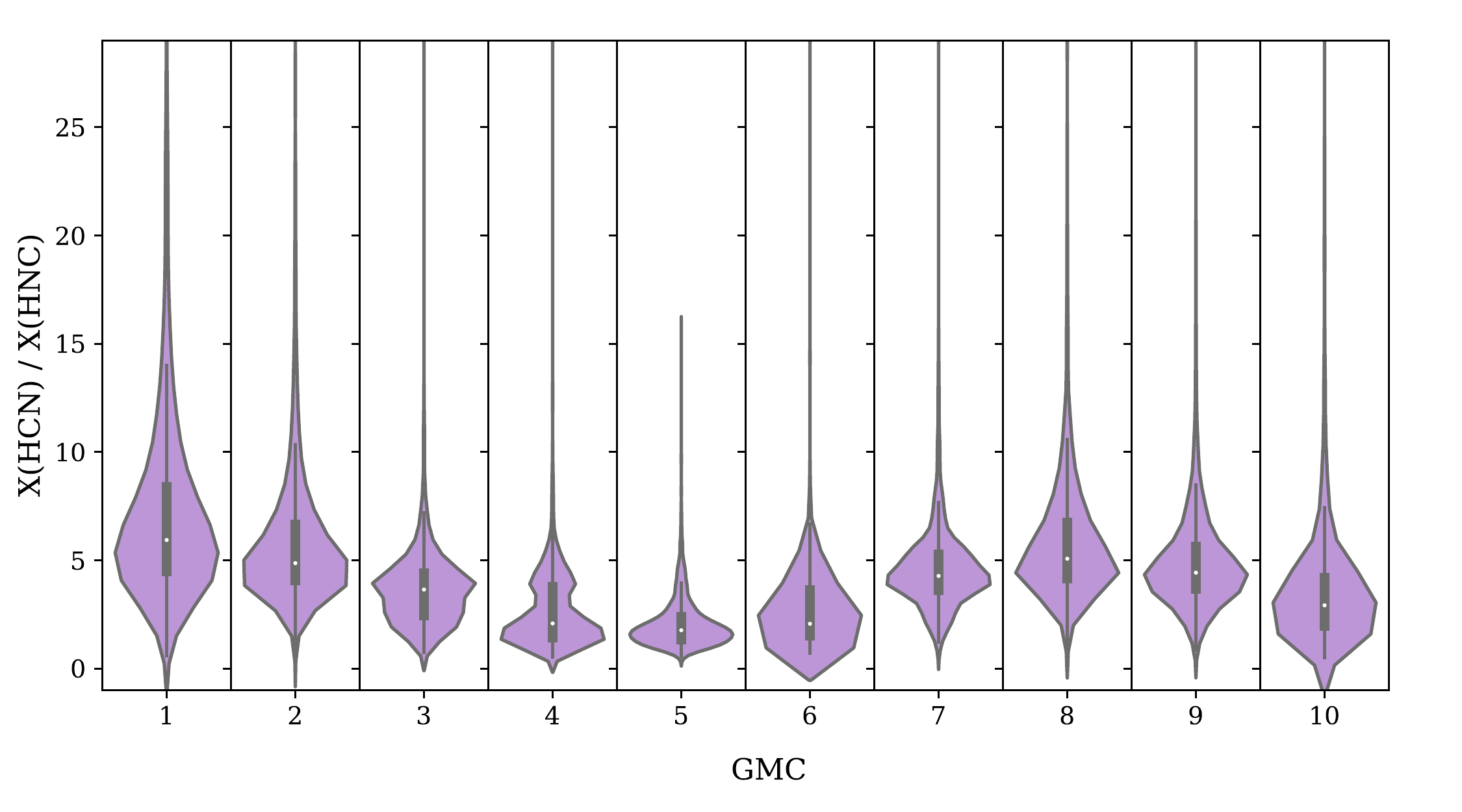}
    \caption{Violin plots derived from modeled HCN/HNC abundance ratios for all GMCs using the most likely 67\% of models. Purple violins indicate the smoothed kernel density estimations of the ratio distributions for each GMC. White dots at the center of each violin indicate the median values of the HCN/HNC ratios derived from \texttt{UCLCHEM}'s abundance estimates. The thick gray vertical bars within each violin show the interquartile ranges of the \texttt{UCLCHEM} datasets, and the thin gray lines illustrate the two outer quartiles.}
    \label{fig:violin_abund}
\end{figure*}

\begin{figure}
    \centering
    \includegraphics[trim=7mm 3mm 5mm 0mm, scale=0.78]{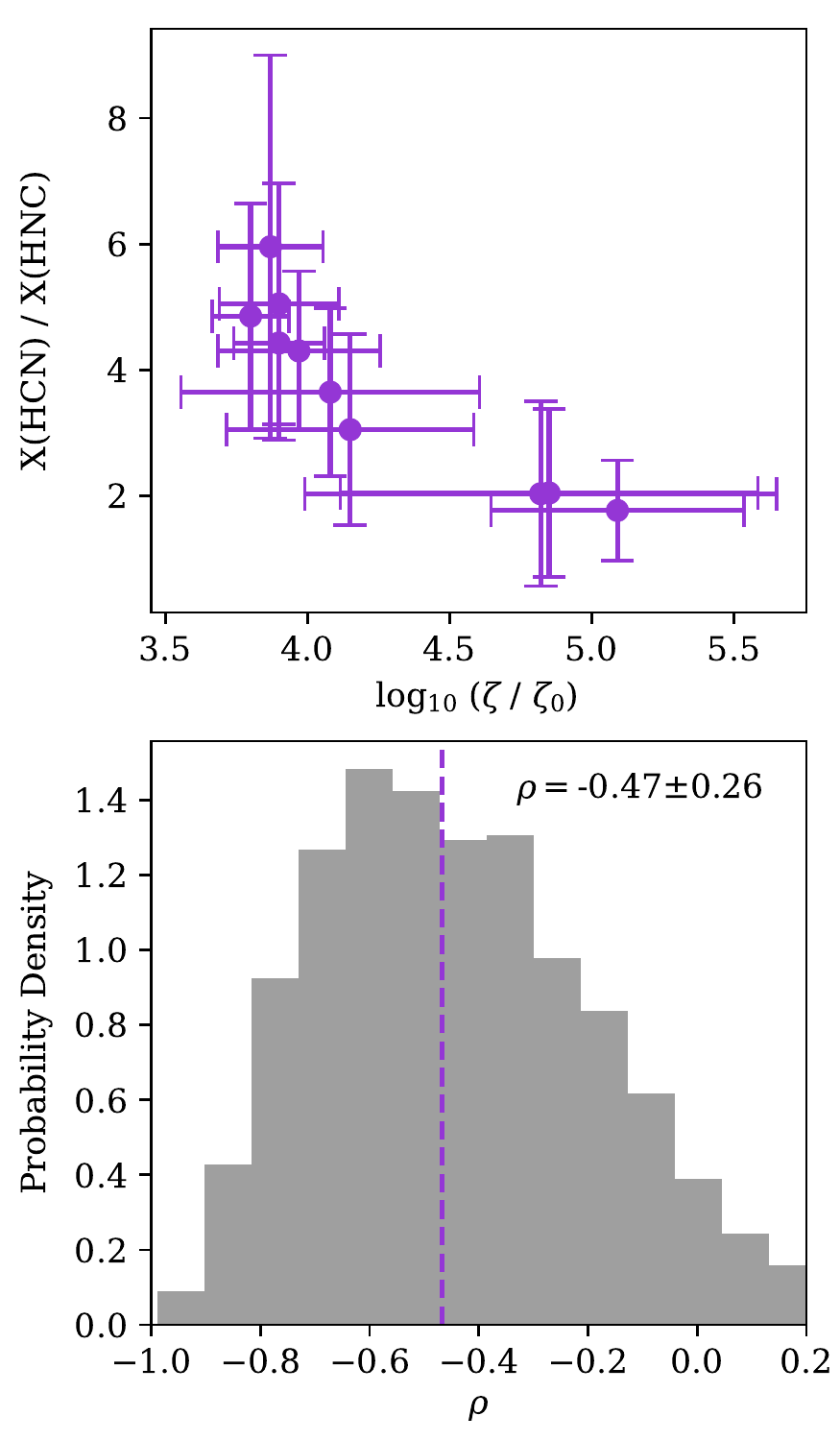}
    \caption{Top: HCN/HNC abundance ratio as a function of cosmic ray ionization rate. Bottom: Distribution of Spearman coefficients for a simulated Gaussian dataset derived from our modeled HCN/HNC abundance ratios, CRIRs, and their uncertainties. The given value of $\rho$ represents the median of the distribution.}
    \label{fig:crir_ratio_corr}
\end{figure}

\begin{figure}
    \centering
    \includegraphics[trim=2mm 3mm 0mm 0mm,scale=0.45]{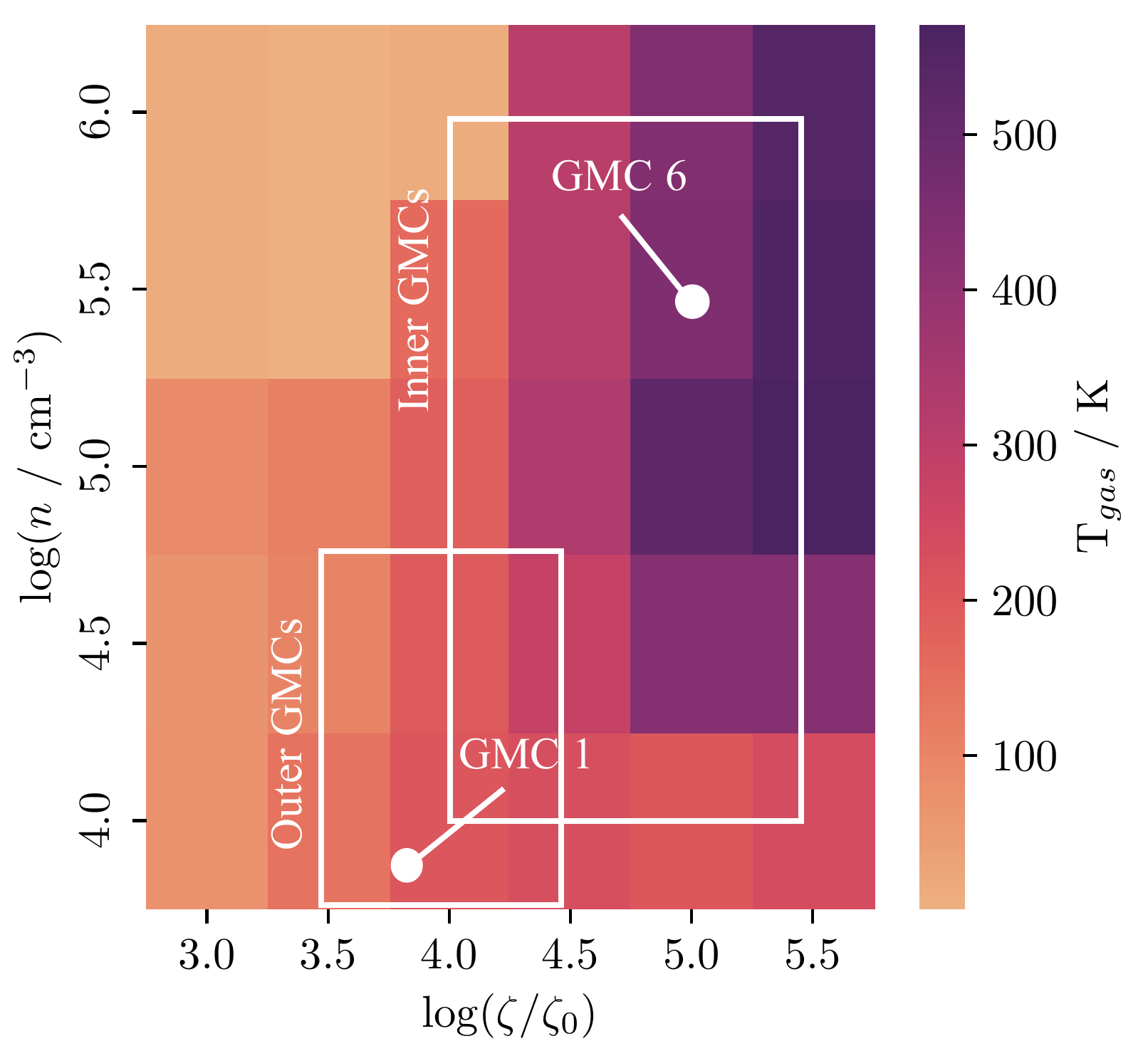}
    \caption{PDR modeling results from \texttt{UCLPDR} for gas temperature as a function of volume density and cosmic ray ionization rate. White boxes indicate the areas of the plot relevant to the conditions in the outer and inner GMCs, and white dots note the specific locations of GMCs 1 and 6.}
    \label{fig:PDR_temp}
\end{figure}

\subsection{Modeling Results} \label{sec:mod_res}

The most likely physical parameters for each GMC as a result of our modeling and sampling algorithms are shown in Table \ref{tab:mod_results} and Figure \ref{fig:samp_results} and are compared in the latter to the number of heating sources per GMC. We show results using the high temperature barriers (2000\,K, 1200\,K) for the HNC + O and HNC + H reactions, as we find varying the temperature barrier had no discernible effect on our results. A possible explanation for this result is presented in Section \ref{sec:cr+ratio}. The most likely parameters found using each of the two temperature barriers were well within the error bars of the opposing model's parameter estimates. In Table \ref{tab:mod_results}, we report the median values of the posterior distributions for each parameter with uncertainties that represent the inner 67\% of the distributions. We find that kinetic temperature and H$_{2}$ column density are largely not constrained by our HCN and HNC measurements.  Nearly all values of kinetic temperature and H$_2$ column density  have an equal likelihood of describing our data, rather than a concentration of points with a high likelihood existing in a small fraction of the parameter space. In cases where the posterior distributions peak at the lower end of our parameter space, we instead report the 83rd percentile of the distribution as an upper limit in Table \ref{tab:mod_results}. However, we are able to constrain volume density and cosmic ray ionization rate, finding $n \sim 10^{4}-10^{5.5}$\,cm$^{-3}$ and $\zeta \sim 10^{-13}-10^{-12}$\,s$^{-1}$ across the GMCs. Our inability to constrain $T_{\text{K}}$ and $N_{\text{H}_{2}}$ is discussed further in Section \ref{sec:cr+ratio}.

We see a bimodality in the CRIR marginalized posterior distributions with solutions at $\sim$\,$10^{4}$\,$\zeta_{0}$ and $\sim$\,$10$\,$\zeta_{0}$ (see Figures \ref{fig:corner1} and \ref{fig:corner6}). We investigated the cause of this bimodality by comparing our model outputs to the data when sampling parameters from each mode. We expected that the low CRIR solution would favor one species or a subset of the transitions, while the high CRIR solution would favor another. However, we find no such physical connection between the low-$\zeta$ solutions and any subset of species or transitions. We suggest that the most likely scenario that can explain the low-$\zeta$ solution is that the part of the parameter space corresponding to low-$\zeta$ values simply happens to produce integrated intensities somewhat close to our measured values. Strong evidence from previous studies based on other sets of molecular lines observed in ALCHEMI \citep{holdship_c2h,harada21,Holdship2022} indicates that the CRIR in the inner GMCs, some of which show bimodality in our models, is $>10^{3}\zeta_0$. As such, we dismiss the lower CRIR solution across all GMCs as unphysical and only present the high CRIR solutions.

To show examples of results for GMCs both in the outer and inner parts of the CMZ, the observed and modeled fluxes for GMCs 1 and 6 are shown in Figure \ref{fig:fluxes}. Observed and modeled fluxes for the remaining GMCs can be found in Appendix \ref{sec:GMCaddcorner}. Figures \ref{fig:corner1} and \ref{fig:corner6} show corner plots for GMCs 1 and 6 respectively, demonstrating our results and the relationships between physical parameters.  Corner plots for GMCs 2--5 and 7--10 can be found in Appendix~\ref{sec:GMCaddcorner}. Our results show an enhancement in volume density and CRIR in the central GMCs (4--6) versus the outer GMCs. One explanation for this increase in volume density and CRIR could be a degeneracy between these two quantities. In Figure \ref{fig:corner6}, the panel showing the relationship between $n$ and $\zeta$ does demonstrate that as $n$ increases by $\sim 2.5$\,dex $\zeta$ increases by $\sim1$\,dex. If $n$ and $\zeta$ were degenerate, we would expect to see an equal change in the spread for both parameters across all GMCs. Thus we believe the behavior demonstrated in Figure \ref{fig:samp_results} is a physical solution rather than resulting from a degeneracy.

Figure \ref{fig:violin_abund} shows a violin plot displaying the HCN/HNC abundance ratios from the best-fitting 67\% of models for all GMCs, where the distribution of values is consistent with the signal-to-noise (S/N) levels in each GMC. GMCs in the center of the nucleus have a higher S/N and exhibit a smaller spread of values when compared to the outer GMCs. We also see that the modeled HCN/HNC abundance ratio is lower in GMCs with higher CRIRs (Figure \ref{fig:crir_ratio_corr}). In order to determine the significance of this apparent anti-correlation, we calculate the Spearman coefficient $\rho$ for the relationship between CRIR and the HCN/HNC abundance ratio by employing the \texttt{SciPy} \citep{scipy2020} function \texttt{spearmanr}\footnote{\url{https://docs.scipy.org/doc/scipy/reference/generated/scipy.stats.spearmanr.html}} \citep{Kokoska1999}. We use our modeled abundance ratios and CRIR estimates for each GMC in combination with their uncertainties to create simulated Gaussian datasets for these parameters consisting of 10,000 sets of 10 data points each (1 point for each GMC)\footnote{\url{https://www.jonathan-liu.com/post/correlationanalysismontecarlo/}}. We calculate $\rho$ for each set of 10 data points in the simulated distributions and find that the median $\rho$ value is $-0.47\pm0.26$ (Figure \ref{fig:crir_ratio_corr}). This $\rho$ value indicates there is a moderately anti-correlated relationship between CRIR and the HCN/HNC abundance ratio.

We find that we are able to reproduce our measured integrated intensity ratios using cosmic ray ionization applied through \texttt{UCLCHEM} gas-grain chemical modeling. Furthermore, the HCN/HNC integrated intensity and abundance ratios are far less than $\sim$50, which was suggested by \cite{meijerink11} for cases of significant mechanical heating. Thus our results place NGC\,253 in the regime of low mechanical heating and high cosmic ray ionization rate, which can be seen in Figure \ref{fig:meijerink_ratio} for the case of no mechanical heating. The HCN and HNC chemistry that results in these low ratios is discussed in Section \ref{sec:cr+ratio}. Our observed integrated intensity ratios are consistent with other extragalactic HCN and HNC measurements (Section \ref{sec:gal_hcn}), suggesting that these galaxies belong to the same part of the heating parameter space as NGC\,253.

Even though our \texttt{RADEX} radiative transfer model accounts for optical depth, there may be a concern that very high optical depths in our HCN and HNC transitions might influence our modeling results. To this end we have used our \texttt{RADEX} analysis to estimate the optical depth of our HCN and HNC transitions. We model the optical depth for all iterations of our \texttt{UltraNest} sampling algorithm and analyze the middle 67\% of the resulting distribution. We find that overall, the optical depth distributions peak at reasonably low ($\lesssim10$) values. We measure slightly higher optical depths in the HCN $1-0$ and $2-1$ transitions, but the optical depths in the corresponding HNC transitions toward all GMCs are similar. The modeled optical depth values across HCN and HNC for all transitions are also close enough in value that we rule out the possibility that different transitions are tracing different physical structures. To confirm these optical depth estimates, we use integrated intensity isotopic ratios for HCN/H$^{13}$CN and HCN/HC$^{15}$N \citep{Martin19}, as well as their isotopomers, to modify our HCN and HNC abundance values from \texttt{UCLCHEM} and thus model the optical depth of these isotopic variants. We found that the ratios of modeled optical depths are consistent with the observationally-derived isotope ratios themselves, confirming that our modeled optical depths are consistent with observations. Additionally, since we model all of the HCN and HNC transitions together, we are effectively deriving an average set of physical parameters across those regions probed by these HCN and HNC transitions. Thus, we conclude that optical depth issues do not inhibit the use of our observed integrated intensities in constraining our models.

\subsection{The Effect of Cosmic Rays on the HCN/HNC Ratio} \label{sec:cr+ratio}

Our results show that the observed HCN/HNC integrated intensity ratios in the NGC\,253 CMZ can be replicated through cosmic ray ionization rates several orders of magnitude higher than those found locally in our own Galaxy. Since \texttt{UCLCHEM}'s treatment of cosmic rays artificially separates cosmic ray ionization from cosmic ray heating, we estimate the heating that would result from these CRIRs via the Photon-Dominated Region (PDR) modeling code \texttt{UCLPDR}\footnote{\url{https://uclchem.github.io/ucl_pdr/}} which treats heating and cooling. \texttt{UCLPDR} provides an estimated gas temperature given input volume densities and CRIRs, assuming an $A_{V} \sim10$ (in this case in order to model the inner, UV-shielded part of the cloud). We see in Figure \ref{fig:PDR_temp} that the density and CRIR estimates in the inner GMCs (4--6) result in PDR gas temperatures of $\sim200-500$\,K, whereas the conditions in the outer GMCs correspond to temperatures of $\sim100-400$\,K. We can assume that if these PDR-derived temperatures were lower than those calculated from observations, the additional heating needed to equate the temperatures would be from mechanical heating. However, the high kinetic temperatures from our PDR modeling agree with those derived by \cite{mangum19} and overlap with the kinetic temperatures found through \texttt{UCLCHEM} modeling, when constrained. Thus we conclude that there is little contribution from mechanical heating and that cosmic ray heating alone can produce high kinetic temperatures. 

Previous studies \citep[\eg][]{Goldsmith1986,Herbst2000,meijerink11,kaz12,Krieger2020ApJ} suggested that the HCN/HNC abundance ratio might probe mechanical heating processes in the ISM through its kinetic temperature sensitivity, with the ratio increasing at high kinetic temperatures. This behavior has been seen in observations of protostellar shocks \citep[\eg][]{Lefloch2021} and C-type shock models, where the HCN/HNC abundance ratio increases by a factor of $\sim$20. Hence we would expect that mechanical heating by shocks in NGC\,253 would produce high HCN/HNC abundance ratios \citep{Viti2017}. Given that we do not observe high ratios in NGC\,253, it appears that shocks are not a dominant source of heating. However, as noted above via PDR modeling, our high CRIRs are still capable of raising the gas temperature, so we must explain how we observe low HCN/HNC ratios while still measuring a high temperature \citep{mangum19}. We therefore investigate the detailed chemical network used by our model and find that identical temperature-independent formation and destruction routes dominate the chemistry of HCN and HNC at high CRIR.

%Despite previous studies \citep[\eg][]{Goldsmith1986,Herbst2000,meijerink11,kaz12,Krieger2020ApJ} which suggested the HCN/HNC ratio was closely tied to mechanical heating as measured through kinetic temperature, the gas temperature is not well constrained by our models indicating that other processes are driving the chemistry. Furthermore, observations of protostellar shocks \citep[\eg][]{Lefloch2021} and C-type shock models have shown that the HCN/HNC abundance ratio increases by a factor of $\sim$ 20 in shocks. Hence we would expect that mechanical heating by shocks in NGC\,253 would produce high HCN/HNC ratios \citep{Viti2017}. Given that we do not observe this and that we find cosmic rays alone are able to heat the gas to these high temperatures, we do not need to invoke mechanical heating to explain the HCN/HNC ratios and the high kinetic temperatures measured in NGC\,253.

%Nevertheless, if mechanical heating produces large HCN/HNC ratios, we must explain how we obtain low ratios even at high temperatures. We therefore investigate the detailed chemical network used by our model and find that identical temperature-independent formation and destruction routes dominate the chemistry of HCN and HNC at high CRIR. 

To see how these chemical pathways drive the HCN and HNC abundance as the CRIR is increased, we start at low CRIR.  The primary source of HCN at low CRIR varies with temperature but is usually a reaction with a small barrier \eg:
\begin{align*}
&{\rm H_2 + CN \longrightarrow HCN + H ~(E_A = 250\,K)} \\
&{\rm N + HCO \longrightarrow HCN + O ~(E_A = 50\,K)} \\
&{\rm N + CH_2 \longrightarrow HCN + H ~(E_A = 50\,K)}
\end{align*}
where E$_A$ is the energy barrier in K.  Routes to form HNC are much less efficient, such that one would expect the HCN/HNC abundance ratio to increase with temperature at low CRIR due to increasing efficiency of HCN formation. This result is consistent with the finding of  \citet[][Figure~\ref{fig:meijerink_ratio}]{meijerink11}, demonstrating that the HCN/HNC abundance ratio is sensitive to and positively correlated with kinetic temperature at low CRIR.

However, at a sufficiently high CRIR, this picture changes. Both species then form mainly through the reactions
\begin{align*}
&{\rm HCNH^+ + e^- \longrightarrow HCN + H}\\
&{\rm HCNH^+ + e^- \longrightarrow HNC + H}
\end{align*}
which have identical rates at all temperatures. Once the CRIR is large enough for these reactions to dominate, both species form at roughly identical rates. Kinetic temperature is no longer as much of an issue, leading to a convergence toward low HCN/HNC abundance ratios across all mechanical heating rates as seen in Figure~\ref{fig:meijerink_ratio}.

The destruction pathways for HCN and HNC are much simpler. Regardless of CRIR, both species are primarily destroyed by reactions with ions. At high CRIR, these proceed much faster because there are more ions but there is no real change of destruction route. In the end, then, the HCN/HNC abundance ratio is largely set by relative formation efficiency rather than destruction. As a result, at high CRIRs, HCN and HNC chemistry is dominated by cosmic rays rather than kinetic temperature. Because NGC\,253 seems to fall in this high end of the CRIR parameter space, our HCN and HNC observations toward NGC\,253 allow us to constrain the cosmic ray ionization rate but not kinetic temperature. Because there is a clear degeneracy between temperature and column density, which is demonstrated by the negative relationship shown in the N$_{\text{H}_{2}}$ versus $T_{\text{K}}$ corner plot panels in Figures \ref{fig:corner1} and \ref{fig:corner6}, the inability to constrain kinetic temperature with our measurements also prevents us from constraining column density.

\section{Discussion} \label{sec:disc}

\subsection{The Influence of Heating Sources}
\label{sec:heatingsources}

A result from our chemical and radiative transfer modeling (Section \ref{sec:model_methods}) of the GMCs in NGC\,253 is an apparent volume density and CRIR gradient in the NGC\,253 CMZ. Figure \ref{fig:samp_results} shows that in GMCs 4, 5, and 6, the predicted density and cosmic ray ionization rate values are upwards of an order of magnitude higher than in the outer GMCs. Our CRIR values agree with those found in other recent molecular studies of the NGC\,253 CMZ. \cite{holdship_c2h} determined CRIRs of 10$^{3}-10^{6}\,\zeta_{0}$ ($\sim10^{-14}-10^{-11}$ s$^{-1}$) could replicate the C$_{2}$H emission seen in the CMZ. Additionally, \cite{harada21} used HOC$^{+}$ observations to estimate $\zeta \gtrsim 10^{-14}$ s$^{-1}$. \cite{Holdship2022} found that H$_{3}$O$^{+}$ and SO measurements corresponded to $\zeta \sim 10^{-13}$ s$^{-1}$, or 10$^{4}\,\zeta_{0}$.

Volume density also appears to be enhanced in GMCs 4, 5, and 6 compared to the outer GMCs. Our density estimates are consistent with those presented in \cite{harada21}, which found $n_{\text{H}} \gtrsim 10^{5}$ cm$^{-3}$ in molecular clumps and $n_{\text{H}} \sim 10^{4.5}$ cm$^{-3}$ in more extended areas of the CMZ. We estimate slightly lower densities than \cite{harada21} in the outer GMCs, with $n \lesssim 10^{4}$ cm$^{-3}$. \cite{leroy15} suggests that the average volume density over the three-dimensional FWHM size of a GMC is $n_{\text{H}_{2}} \sim 2000$ cm$^{-3}$ in these 10 GMCs, which is slightly lower than our estimates.

The highest densities and CRIRs, found in GMCs 4, 5, and 6, are consistent with the density of heating sources present in these clouds. We place heating sources observed using radio continuum \citep{ua97} and vibrationally-excited HC$_{3}$N emission \citep{RV2020} in each of our 10 GMCs by simply identifying which sources fall within a GMC on the plane of the sky (i.e. no distance component is considered). 

We examine the possible relationship between both density and CRIR and the number of heating sources per GMC by calculating Spearman coefficients for the CRIR-heating source and density-heating source relationships. Following the procedure we outlined in Section \ref{sec:mod_res}, we find that the median $\rho$ values for the CRIR-heating source and density-heating source relationships are 0.67 and  0.60 respectively, with standard deviations of 0.21 and 0.17. These values indicate that there are likely positive correlations in the relationships between both CRIR and heating sources as well as density and heating sources. A higher volume density would lead to more favorable star-forming conditions, thus increasing the number of star formation-related heating sources. The increase in the number of heating sources will therefore increase the CRIR, as we expect these heating sources (e.g. supernova remnants) to be the main progenitor of cosmic rays.

The majority of heating sources (HII regions, supernova remnants, and super hot cores\footnote{Keep in mind, though, that the measurements identifying super hot cores \citep{RV2020} sample only the part of the NGC\,253 CMZ encompassing GMCs 3 through 6.}) are located in the nucleus (GMC 5) of the CMZ (Figure~\ref{fig:samp_results}), likely contributing to the enhanced cosmic ray ionization rates predicted there. Though many of these sources are unclassified, we estimate that approximately half of these unclassified sources are supernova remnants producing a high CRIR. This estimate is based on the analysis provided by \cite{ua97}, which determined that 7/14 ($\sigma_\alpha <0.2$) and 8/17 ($\sigma_\alpha < 0.4$) of the sources for which they derived spectral indices had $\alpha \leq -0.4$, indicative of synchrotron emission. Very few heating sources are found in outer GMCs 1, 2, and 8 through 10, which is consistent with our finding that the predicted CRIRs and densities are about an order of magnitude lower than in the nucleus. Furthermore, the sources of the cosmic rays appear to be well correlated with the prevalence of supernovae in the NGC\,253 CMZ.

\subsection{The Connection Between CRIR and Supernovae}
\label{sec:CRIRSNConnection}

In the interest of identifying a possible source for the cosmic rays traced by HCN and HNC chemistry in the NGC\,253 CMZ, we seek to establish a connection between our measured CRIR and supernovae.  In the ISM, the main effect of cosmic rays on ISM chemistry is to initiate and drive the interstellar chemistry by colliding with and ionising atoms and molecules. During ionization they also transfer energy to the ejected electrons and hence heat the gas. While the energies of cosmic rays range from MeV to ultrarelativistic values, the cosmic rays that are primarily responsible for ionizing the ISM are those with energies $\lesssim$ 1 GeV. Measuring the cosmic ray ionization rate below such energies is often done by studying the products of ion-neutral chemistry in the dense ISM.

In our chemical models, $\zeta_{0} = 1.36\times10^{-17}$\,s$^{-1}$ is used as the base CRIR from which all cosmic ray-induced reactions are scaled. This model $\zeta_0$ value appears to be similar to the local Milky Way CRIR.  Analysis by \cite{Webber1998} used data from the \textit{Voyager} and \textit{Pioneer} spacecraft at a distance of 60\,AU from the Sun to estimate the local interstellar cosmic ray spectra and associated energy density and ionization rate lower-limit.  The energy density derived from this analysis is $\sim 1.80$\,eV cm$^{-3}$, while the implied CRIR lower limit is $\zeta_{MW} \gtrsim (3-4)\times10^{-17}$\,s$^{-1}$, within a factor of two of the $\zeta_0$ assumed in our chemical modeling.  Uncertainties in the kinetic energy deposited into the gas per interaction alone (\ie\ the energy produced by ionization of H$_2$ is 20\,eV; \citealt{Goldsmith2001}) are within this range of uncertainty.

It is also important to note that the CRIR in the Milky Way CMZ is measured to be $\sim 1000$ times the local MW CRIR.  \cite{LePetit2016A&A}, using measurements of H$^+_3$, derive CRIRs in the range $1-11\times10^{-14}$\,s$^{-1}$, though this analysis found that this CRIR applies in a medium where the volume density n(H$_2$) $\lesssim 100$\,cm$^{-3}$ to which the H$^+_3$ emission is sensitive.  \cite{Ginsburg2016A&A}, using measurements of the H$_2$CO $3_{03}-2_{02}$ and $3_{21}-2_{20}$ transitions, derive an upper-limit to the CRIR of $\lesssim 10^{-14}$\,s$^{-1}$ in the MW CMZ, constrained by their derived dense gas kinetic temperature of 60\,K.  The H$_2$CO transitions used in this analysis are sensitive to volume densities n$_{\text{H}_{2}}$ $\sim 10^4 - 10^5$\,cm$^{-3}$, similar to the volume densities probed by our HCN and HNC measurements.  \cite{Ginsburg2016A&A} concluded that CR heating is either not dominant in the MW CMZ or is not uniform.

As summarized by \cite{Dalgarno2006}, a lower-limit to the Milky Way CRIR, $\zeta_{MW}$, was established by \cite{Spitzer1968} as $\gtrsim 6.7\times10^{-18}$\,s$^{-1}$ for hydrogen atoms. Also, based on a general consideration of energies released in supernovae, \cite{Spitzer1968} estimated that the probable upper limit to $\zeta_{MW}$ is $1.2\times10^{-15}$\,s$^{-1}$. This upper-limit is obtained by assuming that the atoms in Type I supernova shells, which have an energy of 2\,MeV per nucleon at a velocity of 20,000\,km s$^{-1}$, permeate the Galaxy. If at most one-third of the shell energy of $10^{51}$\,ergs is available to the expanding shell of gas, with an energy loss of 36\,eV per free electron produced during the ionization process, a galactic frequency of one Type I supernova per 100 years gives the upper-limit to $\zeta_{MW}$ quoted.

These analyses suggest a quantitative connection between supernovae and the CRIR where a value for $\zeta_{MW}$ of $1.2\times10^{-15}$\,s$^{-1}$ corresponds roughly to a supernova rate of 0.01\,yr$^{-1}$. Since our CRIR scaling constant is $\zeta_0 = 1.36\times10^{-17}$\,s$^{-1}$, the CRIR represented by $\zeta_0$ corresponds to a supernova rate of $\sim 10^{-4}$\,yr$^{-1}$. The supernova rate for NGC\,253 has been estimated to be in the range 0.14 to 0.3\,yr$^{-1}$ \citep{Lenc2006AJ,ua97}, and an upper limit to the supernova rate of 0.3\,yr$^{-1}$ would imply an upper limit to the CRIR of $\sim 3000$\,$\zeta_0$. This CRIR is on the low end of the range of $\zeta$ values  that we measure toward the GMCs of NGC\,253 (Figure~\ref{fig:samp_results}). The distribution of radio sources with supernova-like spectral indices (Section~\ref{sec:heatingsources}) indicates a higher CRIR within GMCs associated with larger numbers of supernovae in NGC\,253, consistent with the observed trend in CRIR within the CMZ (Figure~\ref{fig:samp_results}).

\section{Conclusions} \label{sec:conc}

We study HCN and HNC emission and its utility in investigating heating processes associated with star formation in the CMZ of the nearby starburst galaxy NGC\,253. Previous studies suggested that the HCN/HNC line ratio would be useful in probing mechanical heating, which was thought to be an abundant heating source in the NGC\,253 CMZ. However, our observations of low HCN/HNC integrated intensity ratios in combination with high kinetic temperatures indicate that either this ratio does not provide insight into the mechanical heating input or that mechanical heating is not a significant heating mechanism in this environment. To understand the implications of our observed integrated intensities, we model the physical conditions in the NGC\,253 CMZ using chemical modeling via \texttt{UCLCHEM} and non-LTE radiative transfer modeling with \texttt{RADEX}. After constraining these models with our HCN and HNC measurements, we come to the following conclusions:

\begin{enumerate}
    \item The HCN/HNC abundance ratios are low ($<10$) in the NGC\,253 CMZ.  This result is consistent with findings in other extragalactic systems but is at odds with previous theoretical work that suggested this ratio should be high ($\gtrsim 50$) in starburst galaxies with substantial mechanical heating \citep{meijerink11,kaz12}.
    \item The HCN/HNC abundance ratios are lowest in GMCs with the highest modeled CRIRs and densities, and we find a moderate anti-correlation between the CRIR and the HCN/HNC ratio (Figure \ref{fig:crir_ratio_corr}).
    \item We see higher cosmic ray ionization rates in the center of the CMZ ($\zeta \sim 10^{-12}$\,s$^{-1}$) as compared to those on its outskirts ($\zeta \sim 10^{-13}$\,s$^{-1}$) (Figure \ref{fig:samp_results}).
    \item Volume density is also enhanced in the central GMCs ($n_{\text{H}_2} \sim 10^{5.5}$\,cm$^{-3}$) as compared to the outer GMCs ($n_{\text{H}_2} \lesssim 10^{4}$\,cm$^{-3}$).
    \item The central GMCs with the highest estimated density and CRIRs also contain the greatest number of heating sources (HII regions, supernova remnants, and super hot cores; Figure \ref{fig:samp_results}) per GMC, with statistical tests indicating a positive correlation between both of these parameters and the number of heating sources per GMC.
    \item Our analysis suggests a quantitative connection between supernovae and the CRIR in NGC\,253. With an estimated supernova rate in the range 0.14 to 0.3\,yr$^{-1}$ \citep{Lenc2006AJ,ua97}, an upper limit to the supernova rate of 0.3\,yr$^{-1}$ would imply an upper limit to the CRIR of $\sim 3000$\,$\zeta_0$. This CRIR is on the low end of the range of CRIRs that we measure toward the GMCs of NGC\,253 (Figure~\ref{fig:samp_results}).
\end{enumerate}

Further work is needed to test the effectiveness of these molecular tracers on other star-forming environments, as different interpretations of the HCN/HNC abundance ratio are possible in cosmic ecosystems exhibiting different conditions. However, NGC\,253 remains an excellent laboratory for studying extragalactic star formation due to its location in our proverbial backyard, and future studies will lay the groundwork for expanding analysis to other galaxies. We hope to further unravel NGC\,253's CMZ by combining tracers from other ALCHEMI studies in order to further constrain the cosmic ray ionization rate, along with other key physical parameters. These data will aid in affirming the interpretations of various tracer molecules and will greatly enhance our understanding of star formation in a starburst environment.
\newpage

\begin{acknowledgments}
We thank the anonymous referee for providing an extremely thorough and constructive review of the original version of this article.  The referee's comments and suggestions resulted in numerous improvements to the research presented in this article, for which we are grateful. We thank Jack Warfield for his technical expertise in getting this project off the ground. We also thank Heihei Behrens for his crucial support to the authors throughout this process. This work is part of a project that has received funding from the European Research Council (ERC) under the European Union’s Horizon 2020 research and innovation programme MOPPEX 833460. V.M.R. acknowledges support from the Comunidad de Madrid through the Atracci\'on de Talento Investigador Modalidad 1 (Doctores con experiencia) Grant (COOL:Cosmic Origins of Life; 2019-T1/TIC-15379). L.C. has received partial support from the Spanish State Research Agency (AEI; project number PID2019-105552RB-C41). N.H. acknowledges support from JSPS KAKENHI Grant Number JP21K03634. PH is a member of and received financial support for this research from the International Max Planck Research School (IMPRS) for Astronomy and Astrophysics at the Universities of Bonn and Cologne. K.S. acknowledges the grant MOST 111-2112-M-001-039 from the Ministry of Science and Technology in Taiwan. 

This paper makes use of the following ALMA data: ADS/JAO.ALMA\#2017.1.00161.L and ADS/JAO.ALMA\#2018.1.00162.S. ALMA is a partnership of ESO (representing its member states), NSF (USA) and NINS (Japan), together with NRC (Canada), MOST and ASIAA (Taiwan), and KASI (Republic of Korea), in cooperation with the Republic of Chile. The Joint ALMA Observatory is operated by ESO, AUI/NRAO and NAOJ.  The National Radio Astronomy Observatory is a facility of the National Science Foundation operated under cooperative agreement by Associated Universities, Inc. 
\end{acknowledgments}

\facility{ALMA}

\software{CASA \citep{casa}, \texttt{Astropy} \citep{Astropy2013}, \texttt{MLFriends} \citep{ultranest14,ultranest19}, \texttt{UltraNest} \citep{ultranest21}, \texttt{SciPy} \citep{scipy2020}}


\newpage
\bibliography{HCN_HNC.bib}{}

\begin{thebibliography}{}
\expandafter\ifx\csname natexlab\endcsname\relax\def\natexlab#1{#1}\fi
\providecommand{\url}[1]{\href{#1}{#1}}
\providecommand{\dodoi}[1]{doi:~\href{http://doi.org/#1}{\nolinkurl{#1}}}
\providecommand{\doeprint}[1]{\href{http://ascl.net/#1}{\nolinkurl{http://ascl.net/#1}}}
\providecommand{\doarXiv}[1]{\href{https://arxiv.org/abs/#1}{\nolinkurl{https://arxiv.org/abs/#1}}}

\bibitem[{{Aalto} {et~al.}(2012){Aalto}, {Garcia-Burillo}, {Muller}, {Winters},
  {van der Werf}, {Henkel}, {Costagliola}, \& {Neri}}]{Aalto2012A&A}
{Aalto}, S., {Garcia-Burillo}, S., {Muller}, S., {et~al.} 2012, \aap, 537, A44,
  \dodoi{10.1051/0004-6361/201117919}

\bibitem[{{Aalto} {et~al.}(2007{\natexlab{a}}){Aalto}, {Monje}, \&
  {Mart{\'\i}n}}]{Aalto2007aA&A}
{Aalto}, S., {Monje}, R., \& {Mart{\'\i}n}, S. 2007{\natexlab{a}}, \aap, 475,
  479, \dodoi{10.1051/0004-6361:20077366}

\bibitem[{{Aalto} {et~al.}(2002){Aalto}, {Polatidis}, {H{\"u}ttemeister}, \&
  {Curran}}]{Aalto2002A&A}
{Aalto}, S., {Polatidis}, A.~G., {H{\"u}ttemeister}, S., \& {Curran}, S.~J.
  2002, \aap, 381, 783, \dodoi{10.1051/0004-6361:20011514}

\bibitem[{{Aalto} {et~al.}(2007{\natexlab{b}}){Aalto}, {Spaans}, {Wiedner}, \&
  {H{\"u}ttemeister}}]{Aalto2007A&A}
{Aalto}, S., {Spaans}, M., {Wiedner}, M.~C., \& {H{\"u}ttemeister}, S.
  2007{\natexlab{b}}, \aap, 464, 193, \dodoi{10.1051/0004-6361:20066473}

\bibitem[{{Aladro} {et~al.}(2015){Aladro}, {Mart{\'\i}n}, {Riquelme}, {Henkel},
  {Mauersberger}, {Mart{\'\i}n-Pintado}, {Wei{\ss}}, {Lefevre}, {Kramer},
  {Requena-Torres}, \& {Armijos-Abenda{\~n}o}}]{Aladro15}
{Aladro}, R., {Mart{\'\i}n}, S., {Riquelme}, D., {et~al.} 2015, \aap, 579,
  A101, \dodoi{10.1051/0004-6361/201424918}

\bibitem[{{Astropy Collaboration} {et~al.}(2013){Astropy Collaboration},
  {Robitaille}, {Tollerud}, {Greenfield}, {Droettboom}, {Bray}, {Aldcroft},
  {Davis}, {Ginsburg}, {Price-Whelan}, {Kerzendorf}, {Conley}, {Crighton},
  {Barbary}, {Muna}, {Ferguson}, {Grollier}, {Parikh}, {Nair}, {Unther},
  {Deil}, {Woillez}, {Conseil}, {Kramer}, {Turner}, {Singer}, {Fox}, {Weaver},
  {Zabalza}, {Edwards}, {Azalee Bostroem}, {Burke}, {Casey}, {Crawford},
  {Dencheva}, {Ely}, {Jenness}, {Labrie}, {Lim}, {Pierfederici}, {Pontzen},
  {Ptak}, {Refsdal}, {Servillat}, \& {Streicher}}]{Astropy2013}
{Astropy Collaboration}, {Robitaille}, T.~P., {Tollerud}, E.~J., {et~al.} 2013,
  \aap, 558, A33, \dodoi{10.1051/0004-6361/201322068}

\bibitem[{{Bayet} {et~al.}(2011){Bayet}, {Williams}, {Hartquist}, \&
  {Viti}}]{bayet11}
{Bayet}, E., {Williams}, D.~A., {Hartquist}, T.~W., \& {Viti}, S. 2011, \mnras,
  414, 1583, \dodoi{10.1111/j.1365-2966.2011.18500.x}

\bibitem[{{Bublitz} {et~al.}(2022){Bublitz}, {Kastner}, {Hily-Blant},
  {Forveille}, {Santander-Garc{\'\i}a}, {Alcolea}, \&
  {Bujarrabal}}]{Bublitz2022A&A}
{Bublitz}, J., {Kastner}, J.~H., {Hily-Blant}, P., {et~al.} 2022, \aap, 659,
  A197, \dodoi{10.1051/0004-6361/202141778}

\bibitem[{{Buchner}(2014)}]{ultranest14}
{Buchner}, J. 2014, Statistics and Computing, 26, 383,
  \dodoi{10.1007/s11222-014-9512-y}

\bibitem[{{Buchner}(2019)}]{ultranest19}
---. 2019, \pasp, 131, 108005, \dodoi{10.1088/1538-3873/aae7fc}

\bibitem[{{Buchner}(2021)}]{ultranest21}
---. 2021, The Journal of Open Source Software, 6, 3001,
  \dodoi{10.21105/joss.03001}

\bibitem[{{Costagliola} {et~al.}(2011){Costagliola}, {Aalto}, {Rodriguez},
  {Muller}, {Spoon}, {Mart{\'\i}n}, {Per{\'e}z-Torres}, {Alberdi}, {Lindberg},
  {Batejat}, {J{\"u}tte}, {van der Werf}, \& {Lahuis}}]{Costagliola2011A&A}
{Costagliola}, F., {Aalto}, S., {Rodriguez}, M.~I., {et~al.} 2011, \aap, 528,
  A30, \dodoi{10.1051/0004-6361/201015628}

\bibitem[{{Costagliola} {et~al.}(2015){Costagliola}, {Sakamoto}, {Muller},
  {Mart{\'\i}n}, {Aalto}, {Harada}, {van der Werf}, {Viti}, {Garcia-Burillo},
  \& {Spaans}}]{Costagliola2015A&A}
{Costagliola}, F., {Sakamoto}, K., {Muller}, S., {et~al.} 2015, \aap, 582, A91,
  \dodoi{10.1051/0004-6361/201526256}

\bibitem[{{Dalgarno}(2006)}]{Dalgarno2006}
{Dalgarno}, A. 2006, Proceedings of the National Academy of Science, 103,
  12269, \dodoi{10.1073/pnas.0602117103}

\bibitem[{{Dyson} \& {Williams}(1997)}]{Dyson1997}
{Dyson}, J.~E., \& {Williams}, D.~A. 1997, {The physics of the interstellar
  medium}, \dodoi{10.1201/9780585368115}

\bibitem[{{Ginsburg} {et~al.}(2016){Ginsburg}, {Henkel}, {Ao}, {Riquelme},
  {Kauffmann}, {Pillai}, {Mills}, {Requena-Torres}, {Immer}, {Testi}, {Ott},
  {Bally}, {Battersby}, {Darling}, {Aalto}, {Stanke}, {Kendrew}, {Kruijssen},
  {Longmore}, {Dale}, {Guesten}, \& {Menten}}]{Ginsburg2016A&A}
{Ginsburg}, A., {Henkel}, C., {Ao}, Y., {et~al.} 2016, \aap, 586, A50,
  \dodoi{10.1051/0004-6361/201526100}

\bibitem[{{Goldsmith}(2001)}]{Goldsmith2001}
{Goldsmith}, P.~F. 2001, \apj, 557, 736, \dodoi{10.1086/322255}

\bibitem[{{Goldsmith} {et~al.}(1986){Goldsmith}, {Irvine}, {Hjalmarson}, \&
  {Ellder}}]{Goldsmith1986}
{Goldsmith}, P.~F., {Irvine}, W.~M., {Hjalmarson}, A., \& {Ellder}, J. 1986,
  ApJ, 310, 383, \dodoi{10.1086/164692}

\bibitem[{{Goldsmith} \& {Kauffmann}(2017)}]{Goldsmith2017ApJ}
{Goldsmith}, P.~F., \& {Kauffmann}, J. 2017, \apj, 841, 25,
  \dodoi{10.3847/1538-4357/aa6f12}

\bibitem[{{Graninger} {et~al.}(2014){Graninger}, {Herbst}, {{\"O}berg}, \&
  {Vasyunin}}]{gran14}
{Graninger}, D.~M., {Herbst}, E., {{\"O}berg}, K.~I., \& {Vasyunin}, A.~I.
  2014, \apj, 787, 74, \dodoi{10.1088/0004-637X/787/1/74}

\bibitem[{{Green} {et~al.}(2016){Green}, {Cunningham}, {Green}, {Dawson},
  {Jones}, {L{\'o}pez-S{\'a}nchez}, {Verdes-Montenegro}, {Henkel}, {Baan}, \&
  {Mart{\'\i}n}}]{Green2016MNRAS}
{Green}, C.~E., {Cunningham}, M.~R., {Green}, J.~A., {et~al.} 2016, \mnras,
  457, 2470, \dodoi{10.1093/mnras/stv2984}

\bibitem[{{Greve} {et~al.}(2009){Greve}, {Papadopoulos}, {Gao}, \&
  {Radford}}]{Greve2009ApJ}
{Greve}, T.~R., {Papadopoulos}, P.~P., {Gao}, Y., \& {Radford}, S.~J.~E. 2009,
  \apj, 692, 1432, \dodoi{10.1088/0004-637X/692/2/1432}

\bibitem[{{Haasler} {et~al.}(2022){Haasler}, {Rivilla}, {Mart{\'\i}n},
  {Holdship}, {Viti}, {Harada}, {Mangum}, {Sakamoto}, {Muller}, {Tanaka},
  {Yoshimura}, {Nakanishi}, {Colzi}, {Hunt}, {Emig}, {Aladro}, {Humire},
  {Henkel}, \& {van der Werf}}]{Haasler2022}
{Haasler}, D., {Rivilla}, V.~M., {Mart{\'\i}n}, S., {et~al.} 2022, \aap, 659,
  A158, \dodoi{10.1051/0004-6361/202142032}

\bibitem[{{Hacar} {et~al.}(2020){Hacar}, {Bosman}, \& {van Dishoeck}}]{hacar20}
{Hacar}, A., {Bosman}, A.~D., \& {van Dishoeck}, E.~F. 2020, \aap, 635, A4,
  \dodoi{10.1051/0004-6361/201936516}

\bibitem[{{Harada} {et~al.}(2021){Harada}, {Mart{\'\i}n}, {Mangum}, {Sakamoto},
  {Muller}, {Tanaka}, {Nakanishi}, {Herrero-Illana}, {Yoshimura}, {M{\"u}hle},
  {Aladro}, {Colzi}, {Rivilla}, {Aalto}, {Behrens}, {Henkel}, {Holdship},
  {Humire}, {Meier}, {Nishimura}, {van der Werf}, \& {Viti}}]{harada21}
{Harada}, N., {Mart{\'\i}n}, S., {Mangum}, J.~G., {et~al.} 2021, \apj, 923, 24,
  \dodoi{10.3847/1538-4357/ac26b8}

\bibitem[{{Herbst} {et~al.}(2000){Herbst}, {Terzieva}, \& {Talbi}}]{Herbst2000}
{Herbst}, E., {Terzieva}, R., \& {Talbi}, D. 2000, MNRAS, 311, 869,
  \dodoi{10.1046/j.1365-8711.2000.03103.x}

\bibitem[{{Hirota} {et~al.}(1998){Hirota}, {Yamamoto}, {Mikami}, \&
  {Ohishi}}]{hirota98}
{Hirota}, T., {Yamamoto}, S., {Mikami}, H., \& {Ohishi}, M. 1998, \apj, 503,
  717, \dodoi{10.1086/306032}

\bibitem[{{Holdship} {et~al.}(2017){Holdship}, {Viti}, {Jim{\'e}nez-Serra},
  {Makrymallis}, \& {Priestley}}]{uclchem}
{Holdship}, J., {Viti}, S., {Jim{\'e}nez-Serra}, I., {Makrymallis}, A., \&
  {Priestley}, F. 2017, \aj, 154, 38, \dodoi{10.3847/1538-3881/aa773f}

\bibitem[{{Holdship} {et~al.}(2021){Holdship}, {Viti}, {Mart{\'\i}n}, {Harada},
  {Mangum}, {Sakamoto}, {Muller}, {Tanaka}, {Yoshimura}, {Nakanishi},
  {Herrero-Illana}, {M{\"u}hle}, {Aladro}, {Colzi}, {Emig},
  {Garc{\'\i}a-Burillo}, {Henkel}, {Humire}, {Meier}, {Rivilla}, \& {van der
  Werf}}]{holdship_c2h}
{Holdship}, J., {Viti}, S., {Mart{\'\i}n}, S., {et~al.} 2021, \aap, 654, A55,
  \dodoi{10.1051/0004-6361/202141233}

\bibitem[{{Holdship} {et~al.}(2022){Holdship}, {Mangum}, {Viti}, {Behrens},
  {Harada}, {Mart{\'\i}n}, {Sakamoto}, {Muller}, {Tanaka}, {Nakanishi},
  {Herrero-Illana}, {Yoshimura}, {Aladro}, {Colzi}, {Emig}, {Henkel},
  {Nishimura}, {Rivilla}, {van der Werf}, \& {Alma Comprehensive
  High-Resolution Extragalactic Molecular Inventory (Alchemi)
  Collaboration}}]{Holdship2022}
{Holdship}, J., {Mangum}, J.~G., {Viti}, S., {et~al.} 2022, ApJ, 931, 89,
  \dodoi{10.3847/1538-4357/ac6753}

\bibitem[{{Humire} {et~al.}(2022){Humire}, {Henkel}, {Hern{\'a}ndez-G{\'o}mez},
  {Mart{\'\i}n}, {Mangum}, {Harada}, {Muller}, {Sakamoto}, {Tanaka},
  {Yoshimura}, {Nakanishi}, {M{\"u}hle}, {Herrero-Illana}, {Meier}, {Caux},
  {Aladro}, {Mauersberger}, {Viti}, {Colzi}, {Rivilla}, {Gorski}, {Menten},
  {Huang}, {Aalto}, {van der Werf}, \& {Emig}}]{Humire2022}
{Humire}, P.~K., {Henkel}, C., {Hern{\'a}ndez-G{\'o}mez}, A., {et~al.} 2022,
  \aap, 663, A33, \dodoi{10.1051/0004-6361/202243384}

\bibitem[{{Imanishi} \& {Nakanishi}(2013)}]{Imanishi2013AJ}
{Imanishi}, M., \& {Nakanishi}, K. 2013, \aj, 146, 91,
  \dodoi{10.1088/0004-6256/146/4/91}

\bibitem[{{Irvine} \& {Schloerb}(1984)}]{irvine84}
{Irvine}, W.~M., \& {Schloerb}, F.~P. 1984, \apj, 282, 516,
  \dodoi{10.1086/162229}

\bibitem[{{Jenkins}(2009)}]{jenkins2009}
{Jenkins}, E.~B. 2009, \apj, 700, 1299, \dodoi{10.1088/0004-637X/700/2/1299}

\bibitem[{{Kamenetzky} {et~al.}(2011){Kamenetzky}, {Glenn}, {Maloney},
  {Aguirre}, {Bock}, {Bradford}, {Earle}, {Inami}, {Matsuhara}, {Murphy},
  {Naylor}, {Nguyen}, \& {Zmuidzinas}}]{Kamenetzky2011ApJ}
{Kamenetzky}, J., {Glenn}, J., {Maloney}, P.~R., {et~al.} 2011, \apj, 731, 83,
  \dodoi{10.1088/0004-637X/731/2/83}

\bibitem[{{Kazandjian} {et~al.}(2012){Kazandjian}, {Meijerink}, {Pelupessy},
  {Israel}, \& {Spaans}}]{kaz12}
{Kazandjian}, M.~V., {Meijerink}, R., {Pelupessy}, I., {Israel}, F.~P., \&
  {Spaans}, M. 2012, \aap, 542, A65, \dodoi{10.1051/0004-6361/201118641}

\bibitem[{Kokoska \& Zwillinger(1999)}]{Kokoska1999}
Kokoska, S., \& Zwillinger, D. 1999, in CRC Standard Probability and Statistics
  Tables and Formulae, Student Edition

\bibitem[{{Krieger} {et~al.}(2020){Krieger}, {Bolatto}, {Leroy}, {Levy},
  {Mills}, {Meier}, {Ott}, {Veilleux}, {Walter}, \&
  {Wei{\ss}}}]{Krieger2020ApJ}
{Krieger}, N., {Bolatto}, A.~D., {Leroy}, A.~K., {et~al.} 2020, \apj, 897, 176,
  \dodoi{10.3847/1538-4357/ab9c23}

\bibitem[{{Le Petit} {et~al.}(2016){Le Petit}, {Ruaud}, {Bron}, {Godard},
  {Roueff}, {Languignon}, \& {Le Bourlot}}]{LePetit2016A&A}
{Le Petit}, F., {Ruaud}, M., {Bron}, E., {et~al.} 2016, \aap, 585, A105,
  \dodoi{10.1051/0004-6361/201526658}

\bibitem[{{Lefloch} {et~al.}(2021){Lefloch}, {Busquet}, {Viti}, {Vastel},
  {Mendoza}, {Benedettini}, {Codella}, {Podio}, {Schutzer}, {Rivera-Ortiz},
  {L{\'e}pine}, \& {Bachiller}}]{Lefloch2021}
{Lefloch}, B., {Busquet}, G., {Viti}, S., {et~al.} 2021, \mnras, 507, 1034,
  \dodoi{10.1093/mnras/stab2134}

\bibitem[{{Lenc} \& {Tingay}(2006)}]{Lenc2006AJ}
{Lenc}, E., \& {Tingay}, S.~J. 2006, \aj, 132, 1333, \dodoi{10.1086/506475}

\bibitem[{{Leroy} {et~al.}(2015){Leroy}, {Bolatto}, {Ostriker}, {Rosolowsky},
  {Walter}, {Warren}, {Donovan Meyer}, {Hodge}, {Meier}, {Ott}, {Sandstrom},
  {Schruba}, {Veilleux}, \& {Zwaan}}]{leroy15}
{Leroy}, A.~K., {Bolatto}, A.~D., {Ostriker}, E.~C., {et~al.} 2015, \apj, 801,
  25, \dodoi{10.1088/0004-637X/801/1/25}

\bibitem[{{Leroy} {et~al.}(2018){Leroy}, {Bolatto}, {Ostriker}, {Walter},
  {Gorski}, {Ginsburg}, {Krieger}, {Levy}, {Meier}, {Mills}, {Ott},
  {Rosolowsky}, {Thompson}, {Veilleux}, \& {Zschaechner}}]{leroy18}
---. 2018, \apj, 869, 126, \dodoi{10.3847/1538-4357/aaecd1}

\bibitem[{{Li} {et~al.}(2021){Li}, {Wang}, {Gao}, {Liu}, {Zhang}, {Li}, {Gong},
  {Li}, \& {Shi}}]{Li2021MNRAS}
{Li}, F., {Wang}, J., {Gao}, F., {et~al.} 2021, \mnras, 503, 4508,
  \dodoi{10.1093/mnras/stab745}

\bibitem[{{Mangum} {et~al.}(2019){Mangum}, {Ginsburg}, {Henkel}, {Menten},
  {Aalto}, \& {van der Werf}}]{mangum19}
{Mangum}, J.~G., {Ginsburg}, A.~G., {Henkel}, C., {et~al.} 2019, \apj, 871,
  170, \dodoi{10.3847/1538-4357/aafa15}

\bibitem[{{Mart{\'\i}n} {et~al.}(2019){Mart{\'\i}n}, {Muller}, {Henkel},
  {Meier}, {Aladro}, {Sakamoto}, \& {van der Werf}}]{Martin19}
{Mart{\'\i}n}, S., {Muller}, S., {Henkel}, C., {et~al.} 2019, \aap, 624, A125,
  \dodoi{10.1051/0004-6361/201935106}

\bibitem[{{Mart{\'\i}n} {et~al.}(2021){Mart{\'\i}n}, {Mangum}, {Harada},
  {Costagliola}, {Sakamoto}, {Muller}, {Aladro}, {Tanaka}, {Yoshimura},
  {Nakanishi}, {Herrero-Illana}, {M{\"u}hle}, {Aalto}, {Behrens}, {Colzi},
  {Emig}, {Fuller}, {Garc{\'\i}a-Burillo}, {Greve}, {Henkel}, {Holdship},
  {Humire}, {Hunt}, {Izumi}, {Kohno}, {K{\"o}nig}, {Meier}, {Nakajima},
  {Nishimura}, {Padovani}, {Rivilla}, {Takano}, {van der Werf}, {Viti}, \&
  {Yan}}]{ALCHEMI-ACA}
{Mart{\'\i}n}, S., {Mangum}, J.~G., {Harada}, N., {et~al.} 2021, \aap, 656,
  A46, \dodoi{10.1051/0004-6361/202141567}

\bibitem[{{Mauersberger} {et~al.}(2003){Mauersberger}, {Henkel}, {Wei{\ss}},
  {Peck}, \& {Hagiwara}}]{Mauersberger2003}
{Mauersberger}, R., {Henkel}, C., {Wei{\ss}}, A., {Peck}, A.~B., \& {Hagiwara},
  Y. 2003, A\&A, 403, 561, \dodoi{10.1051/0004-6361:20030386}

\bibitem[{{McCormick} {et~al.}(2013){McCormick}, {Veilleux}, \&
  {Rupke}}]{incl_McCormick}
{McCormick}, A., {Veilleux}, S., \& {Rupke}, D. S.~N. 2013, \apj, 774, 126,
  \dodoi{10.1088/0004-637X/774/2/126}

\bibitem[{{McElroy} {et~al.}(2013){McElroy}, {Walsh}, {Markwick}, {Cordiner},
  {Smith}, \& {Millar}}]{McElroy2013}
{McElroy}, D., {Walsh}, C., {Markwick}, A.~J., {et~al.} 2013, \aap, 550, A36,
  \dodoi{10.1051/0004-6361/201220465}

\bibitem[{{McMullin} {et~al.}(2007){McMullin}, {Waters}, {Schiebel}, {Young},
  \& {Golap}}]{casa}
{McMullin}, J.~P., {Waters}, B., {Schiebel}, D., {Young}, W., \& {Golap}, K.
  2007, in Astronomical Society of the Pacific Conference Series, Vol. 376,
  Astronomical Data Analysis Software and Systems XVI, ed. R.~A. {Shaw},
  F.~{Hill}, \& D.~J. {Bell}, 127

\bibitem[{{Meijerink} {et~al.}(2011){Meijerink}, {Spaans}, {Loenen}, \& {van
  der Werf}}]{meijerink11}
{Meijerink}, R., {Spaans}, M., {Loenen}, A.~F., \& {van der Werf}, P.~P. 2011,
  \aap, 525, A119, \dodoi{10.1051/0004-6361/201015136}

\bibitem[{{M{\"u}ller-S{\'a}nchez} {et~al.}(2010){M{\"u}ller-S{\'a}nchez},
  {Gonz{\'a}lez-Mart{\'\i}n}, {Fern{\'a}ndez-Ontiveros}, {Acosta-Pulido}, \&
  {Prieto}}]{ms10}
{M{\"u}ller-S{\'a}nchez}, F., {Gonz{\'a}lez-Mart{\'\i}n}, O.,
  {Fern{\'a}ndez-Ontiveros}, J.~A., {Acosta-Pulido}, J.~A., \& {Prieto}, M.~A.
  2010, \apj, 716, 1166, \dodoi{10.1088/0004-637X/716/2/1166}

\bibitem[{{Papadopoulos}(2010)}]{papadopoulos}
{Papadopoulos}, P.~P. 2010, \apj, 720, 226, \dodoi{10.1088/0004-637X/720/1/226}

\bibitem[{{P{\'e}rez-Beaupuits} {et~al.}(2007){P{\'e}rez-Beaupuits}, {Aalto},
  \& {Gerebro}}]{Perez-Beaupuits2007A&A}
{P{\'e}rez-Beaupuits}, J.~P., {Aalto}, S., \& {Gerebro}, H. 2007, \aap, 476,
  177, \dodoi{10.1051/0004-6361:20078479}

\bibitem[{{Rekola} {et~al.}(2005){Rekola}, {Richer}, {McCall}, {Valtonen},
  {Kotilainen}, \& {Flynn}}]{dist_Rekola}
{Rekola}, R., {Richer}, M.~G., {McCall}, M.~L., {et~al.} 2005, \mnras, 361,
  330, \dodoi{10.1111/j.1365-2966.2005.09166.x}

\bibitem[{{Rico-Villas} {et~al.}(2020){Rico-Villas}, {Mart{\'\i}n-Pintado},
  {Gonz{\'a}lez-Alfonso}, {Mart{\'\i}n}, \& {Rivilla}}]{RV2020}
{Rico-Villas}, F., {Mart{\'\i}n-Pintado}, J., {Gonz{\'a}lez-Alfonso}, E.,
  {Mart{\'\i}n}, S., \& {Rivilla}, V.~M. 2020, \mnras, 491, 4573,
  \dodoi{10.1093/mnras/stz3347}

\bibitem[{{Schilke} {et~al.}(1992){Schilke}, {Walmsley}, {Pineau Des Forets},
  {Roueff}, {Flower}, \& {Guilloteau}}]{Schilke1992}
{Schilke}, P., {Walmsley}, C.~M., {Pineau Des Forets}, G., {et~al.} 1992, \aap,
  256, 595

\bibitem[{{Spilker} {et~al.}(2014){Spilker}, {Marrone}, {Aguirre}, {Aravena},
  {Ashby}, {B{\'e}thermin}, {Bradford}, {Bothwell}, {Brodwin}, {Carlstrom},
  {Chapman}, {Crawford}, {de Breuck}, {Fassnacht}, {Gonzalez}, {Greve},
  {Gullberg}, {Hezaveh}, {Holzapfel}, {Husband}, {Ma}, {Malkan}, {Murphy},
  {Reichardt}, {Rotermund}, {Stalder}, {Stark}, {Strandet}, {Vieira},
  {Wei{\ss}}, \& {Welikala}}]{Spilker2014ApJ}
{Spilker}, J.~S., {Marrone}, D.~P., {Aguirre}, J.~E., {et~al.} 2014, \apj, 785,
  149, \dodoi{10.1088/0004-637X/785/2/149}

\bibitem[{{Spitzer} \& {Tomasko}(1968)}]{Spitzer1968}
{Spitzer}, Lyman, J., \& {Tomasko}, M.~G. 1968, \apj, 152, 971,
  \dodoi{10.1086/149610}

\bibitem[{{Turner} \& {Ho}(1985{\natexlab{a}})}]{th85}
{Turner}, J.~L., \& {Ho}, P.~T.~P. 1985{\natexlab{a}}, \apjl, 299, L77,
  \dodoi{10.1086/184584}

\bibitem[{{Turner} \& {Ho}(1985{\natexlab{b}})}]{Turner1985}
---. 1985{\natexlab{b}}, \apjl, 299, L77, \dodoi{10.1086/184584}

\bibitem[{{Ulvestad} \& {Antonucci}(1997)}]{ua97}
{Ulvestad}, J.~S., \& {Antonucci}, R. R.~J. 1997, \apj, 488, 621,
  \dodoi{10.1086/304739}

\bibitem[{{van der Tak} {et~al.}(2007){van der Tak}, {Black}, {Sch{\"o}ier},
  {Jansen}, \& {van Dishoeck}}]{radex}
{van der Tak}, F.~F.~S., {Black}, J.~H., {Sch{\"o}ier}, F.~L., {Jansen}, D.~J.,
  \& {van Dishoeck}, E.~F. 2007, \aap, 468, 627,
  \dodoi{10.1051/0004-6361:20066820}

\bibitem[{Virtanen {et~al.}(2020)Virtanen, Gommers, Oliphant, Haberland, Reddy,
  Cournapeau, Burovski, Peterson, Weckesser, Bright, {van der Walt}, Brett,
  Wilson, Millman, Mayorov, Nelson, Jones, Kern, Larson, Carey, Polat, Feng,
  Moore, {VanderPlas}, Laxalde, Perktold, Cimrman, Henriksen, Quintero, Harris,
  Archibald, Ribeiro, Pedregosa, {van Mulbregt}, \& {SciPy 1.0
  Contributors}}]{scipy2020}
Virtanen, P., Gommers, R., Oliphant, T.~E., {et~al.} 2020, Nature Methods, 17,
  261, \dodoi{10.1038/s41592-019-0686-2}

\bibitem[{{Viti}(2017)}]{Viti2017}
{Viti}, S. 2017, \aap, 607, A118, \dodoi{10.1051/0004-6361/201628877}

\bibitem[{{Webber}(1998)}]{Webber1998}
{Webber}, W.~R. 1998, \apj, 506, 329, \dodoi{10.1086/306222}

\end{thebibliography}
\bibliographystyle{aasjournal}

\restartappendixnumbering
\appendix 

\section{GMC Positions}
\label{sec:GMCpos}

The GMC positions we have adopted in this analysis (Table~\ref{tab:GMCpos}) are derived from the GMC positions reported by \cite{leroy15}, which are derived from an analysis of $\sim 1.5$\,arcsec imaging of the HCN, HCO$^+$, and CS $2-1$ emission toward the NGC\,253 CMZ.  The GMC positions listed in Table~\ref{tab:GMCpos} differ from those listed in Table~3 of \cite{leroy15} in two minor ways (Leroy, A.K., private communication):
\begin{itemize}
    \item The reference for the offset positions listed in \cite{leroy15}, Table~3, should be R.A.(J2000) = 00$^h$47$^m$33$^s$.1442, Dec.(J2000) = $-25^\circ$17$^\prime$18$^{\prime\prime}$.0024.
    \item GMC~5 has been shifted down in declination by 0.5\,arcsec relative to that reported in \cite{leroy15}.
\end{itemize}
The resultant differences between the GMC positions reported by \cite{leroy15} and those in Table~\ref{tab:GMCpos} are less than 1.5\,arcsec. The GMC positions listed in Table~\ref{tab:GMCpos} are also within 0.5\,arcsec of the continuum source positions derived from the 218 through 365\,GHz continuum images presented in \cite{mangum19}.

\begin{deluxetable}{ccc}
\centering
\tablecolumns{3}
\tablewidth{0pt}
\tablecaption{NGC\,253 GMC Positions\label{tab:GMCpos}}
\tablehead{
\colhead{GMC} & \colhead{R.A.(ICRS)} & \colhead{Dec.(ICRS)} \\
& \colhead{(00$^h$ 47$^m$)} & \colhead{($-25^\circ$ 17$^\prime$)}
}
\startdata
GMC\,1 & 32$^s$.0184 & 28$^{\prime\prime}$.248 \\
GMC\,2 & 32$^s$.2776 & 20$^{\prime\prime}$.22s \\
GMC\,3 & 32$^s$.8056 & 21$^{\prime\prime}$.552 \\
GMC\,4 & 32$^s$.9736 & 19$^{\prime\prime}$.968 \\
GMC\,5 & 33$^s$.2112 & 17$^{\prime\prime}$.412 \\
GMC\,6 & 33$^s$.3312 & 15$^{\prime\prime}$.756 \\
GMC\,7 & 33$^s$.6432 & 13$^{\prime\prime}$.272 \\
GMC\,8 & 34$^s$.0224 & 11$^{\prime\prime}$.400 \\
GMC\,9 & 34$^s$.1664 & 12$^{\prime\prime}$.264 \\
GMC\,10 & 34$^s$.236 & 07$^{\prime\prime}$.836 \\
\enddata
\end{deluxetable}

\section{GMC 2-5 and 7-10 Model Result Corner Plots}
\label{sec:GMCaddcorner}
In this Appendix we show modeling results for GMC\,2 through 5 and 7 through 10 as corner plots.

\begin{figure*}[htbp]
    \centering
    \includegraphics[scale=0.65]{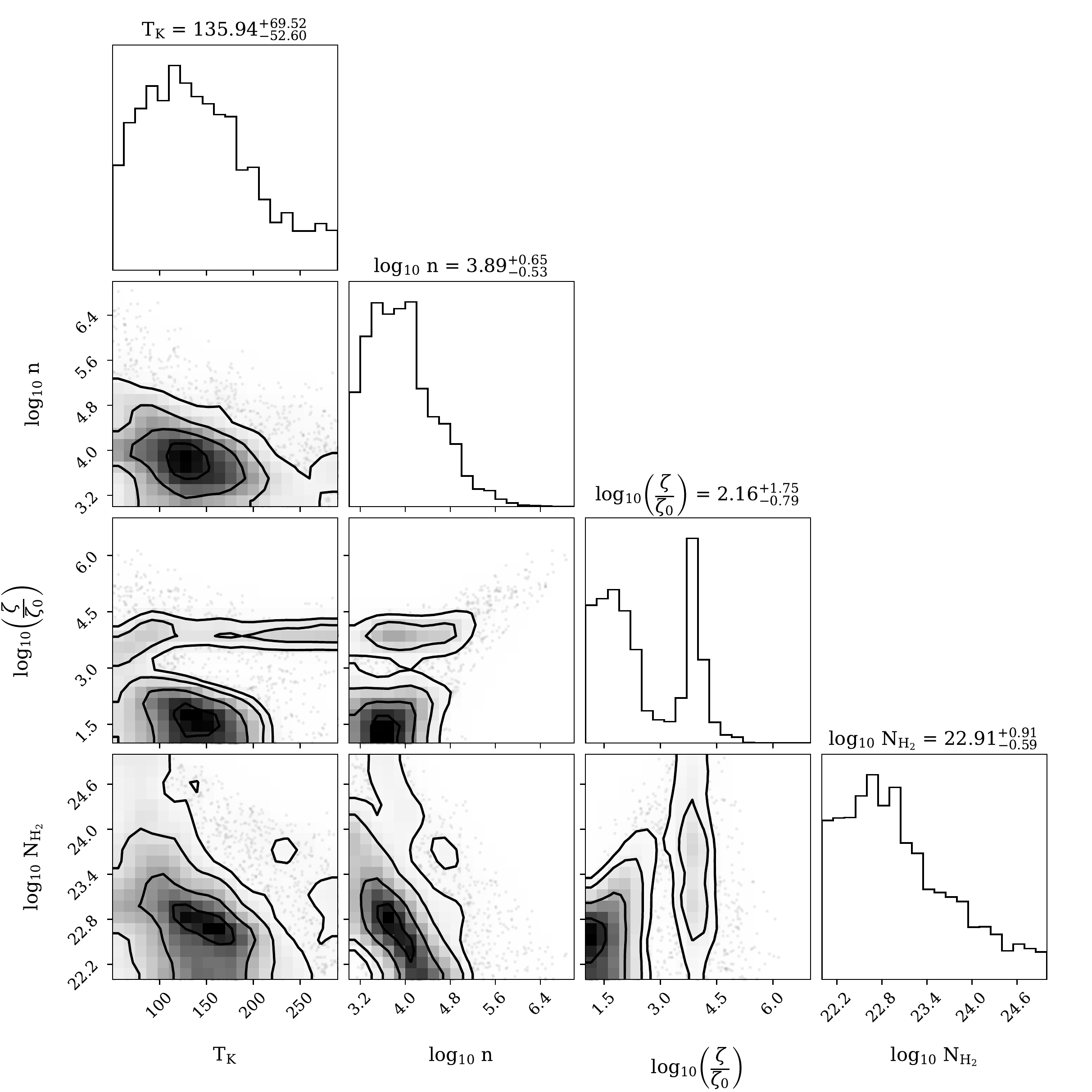}
    \caption{Modeling results for GMC 2.}
    \label{fig:corner2}
\end{figure*}

\begin{figure*}[htbp]
    \centering
    \includegraphics[scale=0.65]{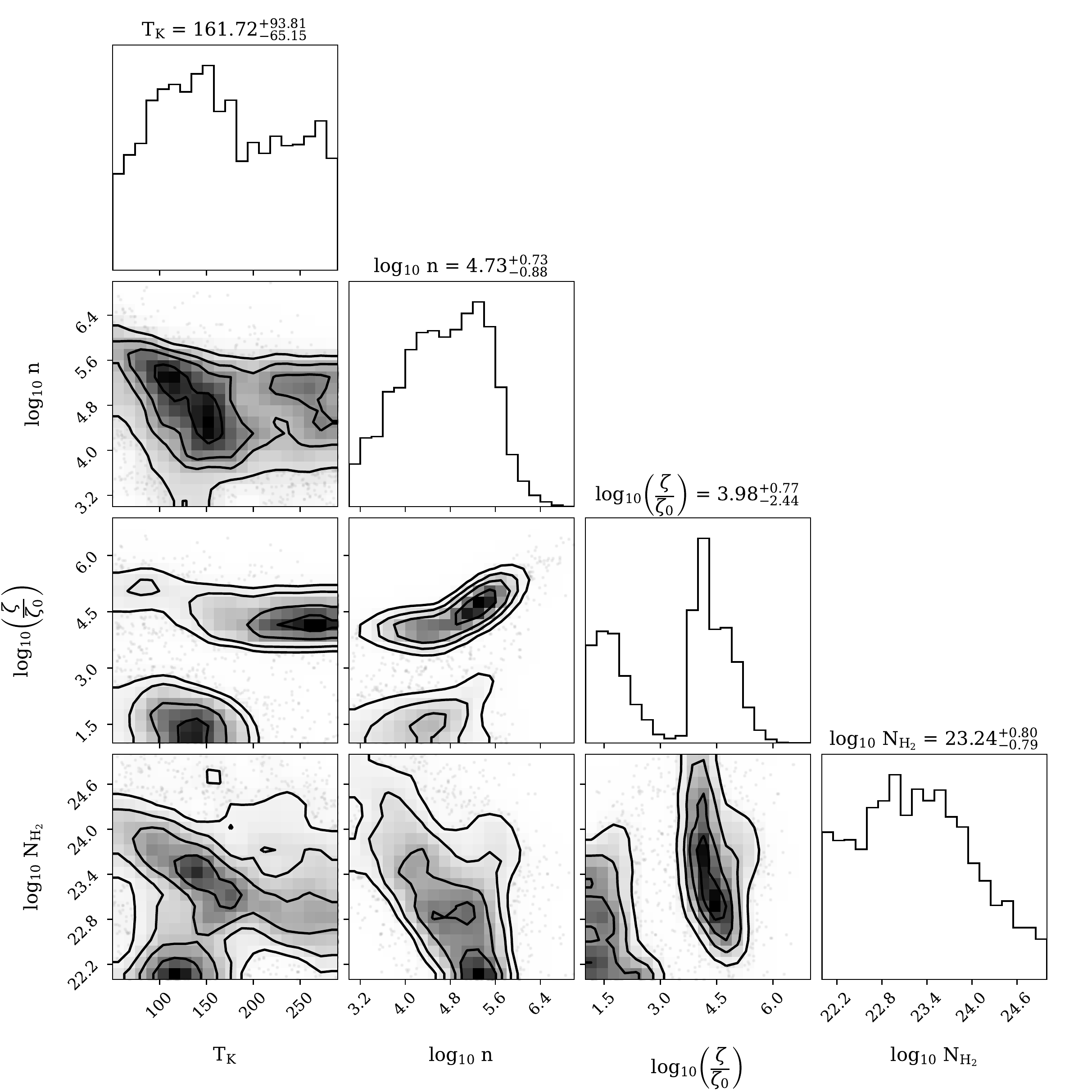}  
    \caption{Modeling results for GMC 3.}
    \label{fig:corner3}
\end{figure*}

\begin{figure*}[htbp]
    \centering
    \includegraphics[trim = 3mm 0mm 0mm 0mm, scale=0.65]{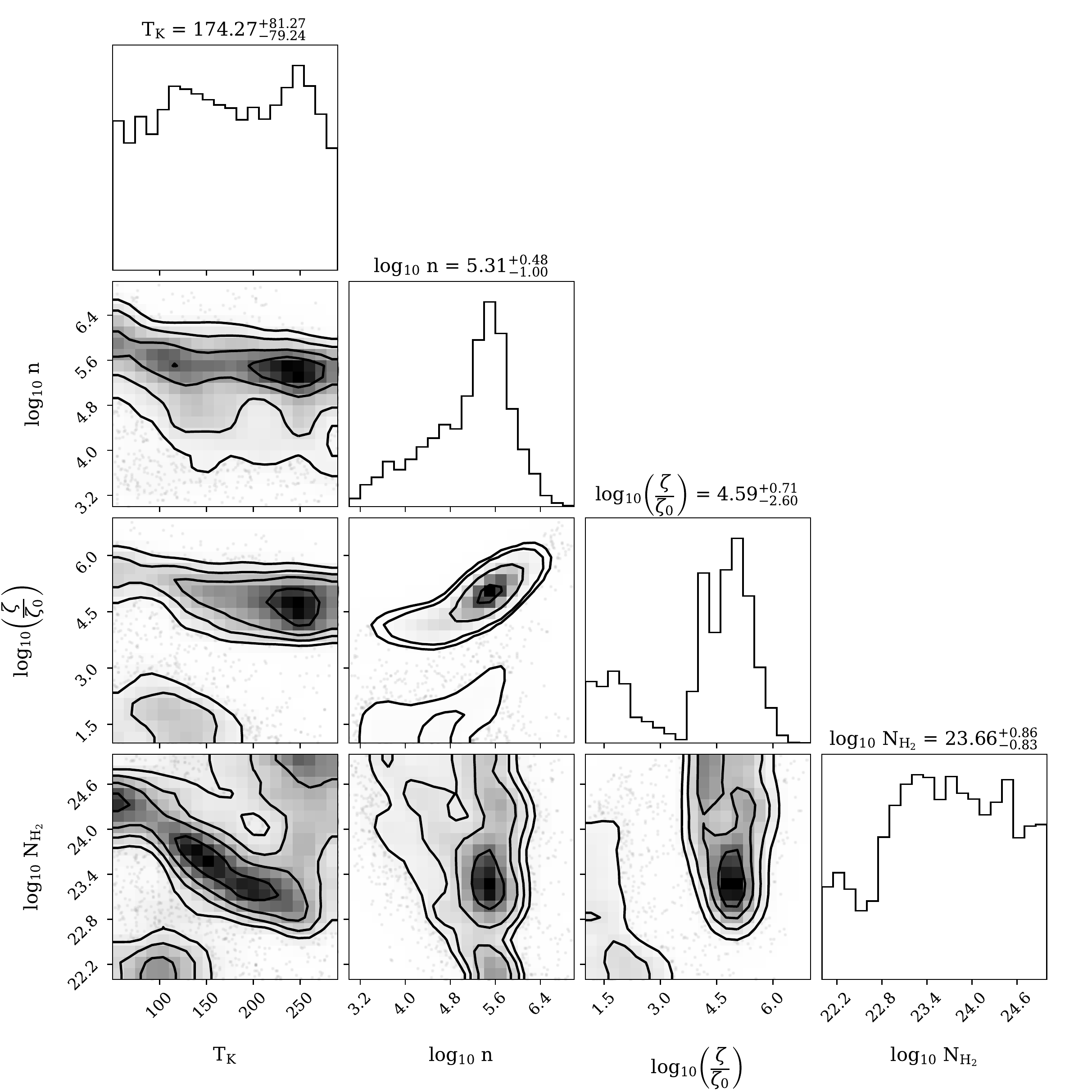}
    \caption{Modeling results for GMC 4.}
    \label{fig:corner4}
\end{figure*}

\begin{figure*}[htbp]
    \centering
    \includegraphics[scale=0.65]{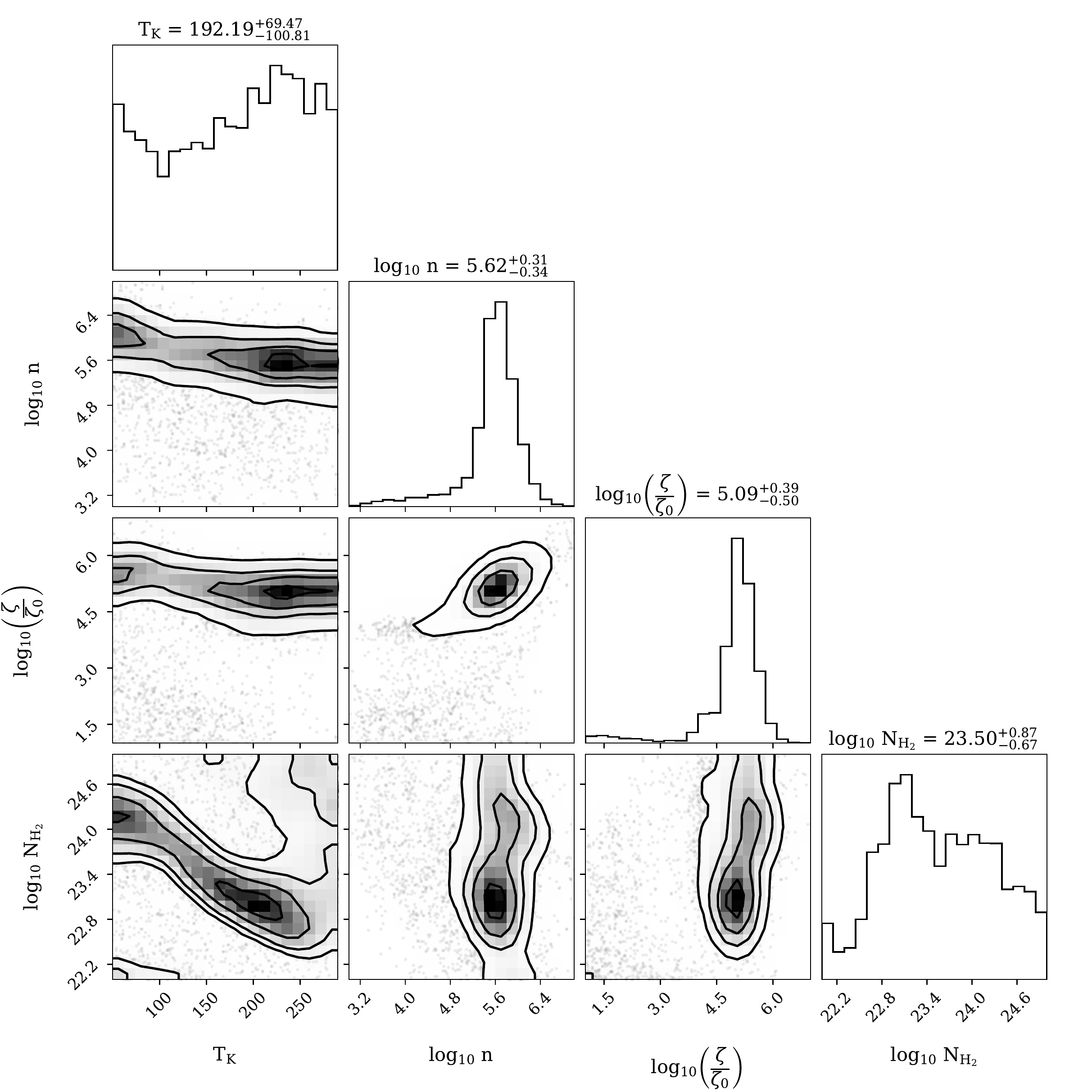}
    \caption{Modeling results for GMC 5.}
    \label{fig:corner5}
\end{figure*}

\begin{figure*}[htbp]
    \centering
    \includegraphics[scale=0.65]{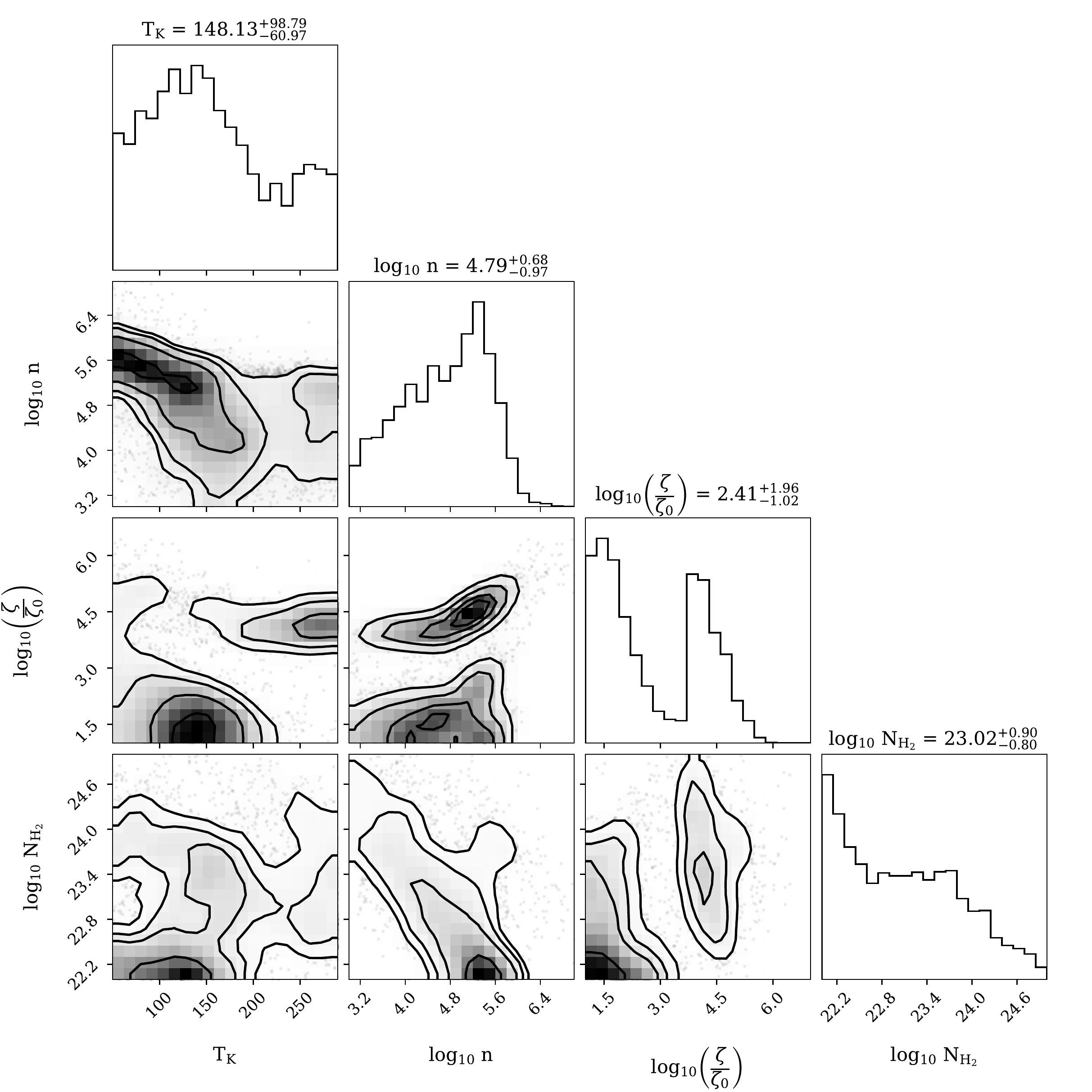}
    \caption{Modeling results for GMC 7.}
    \label{fig:corner7}
\end{figure*}

\begin{figure*}[htbp]
    \centering
    \includegraphics[scale=0.65]{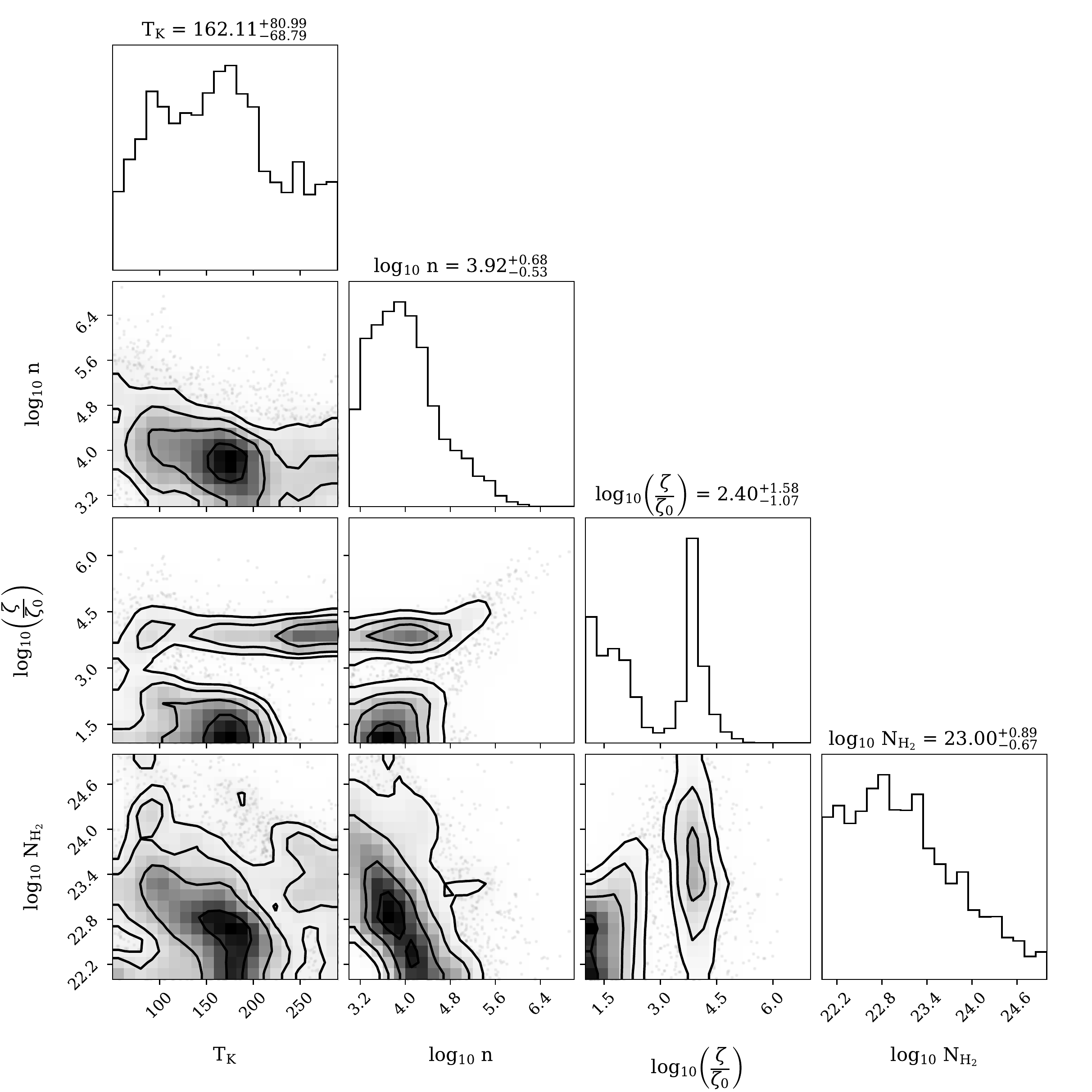}
    \caption{Modeling results for GMC 8.}
    \label{fig:corner8}
\end{figure*}

\begin{figure*}[htbp]
    \centering
    \includegraphics[scale=0.65]{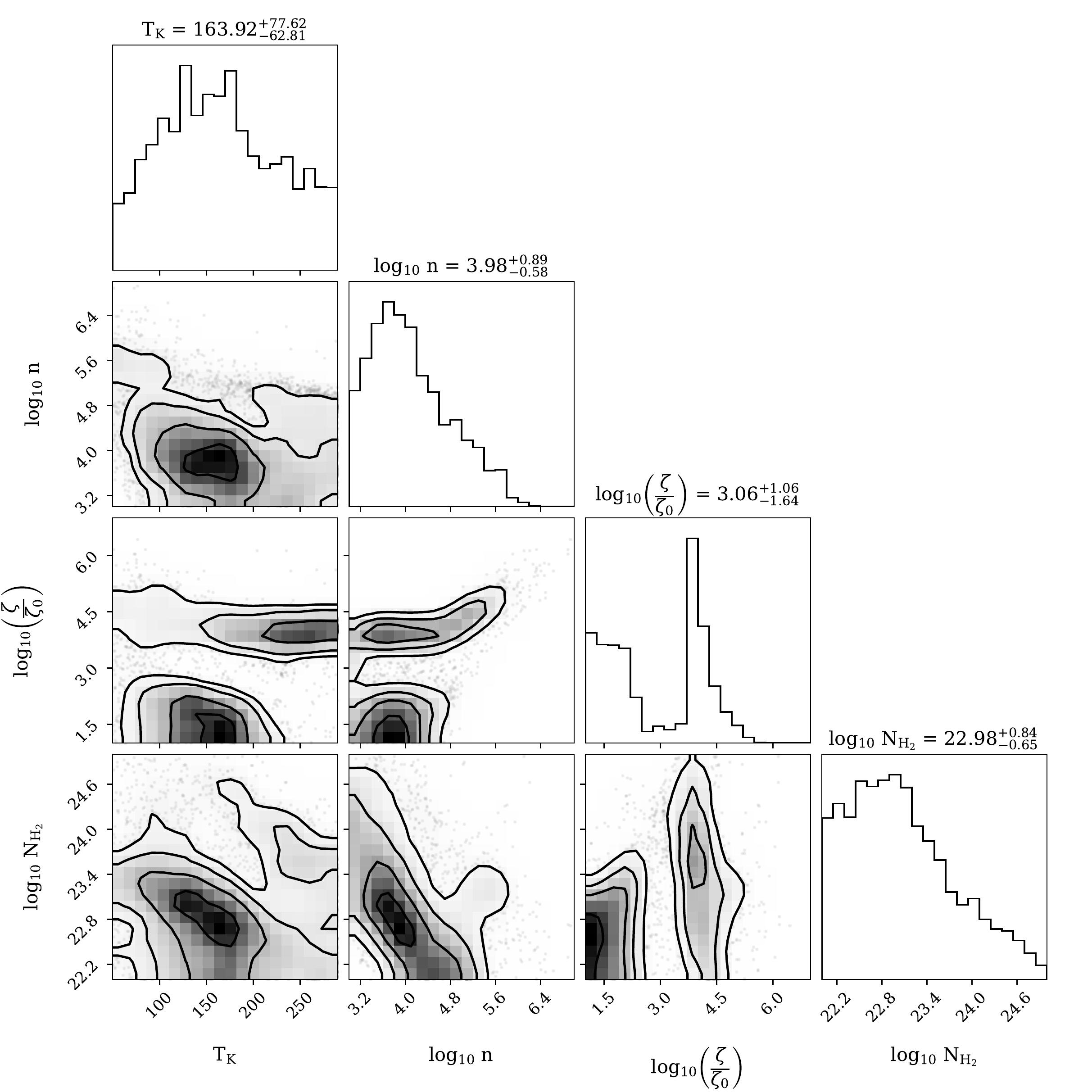}
    \caption{Modeling results for GMC 9.}
    \label{fig:corner9}
\end{figure*}

\begin{figure*}[htbp]
    \centering
    \includegraphics[scale=0.65]{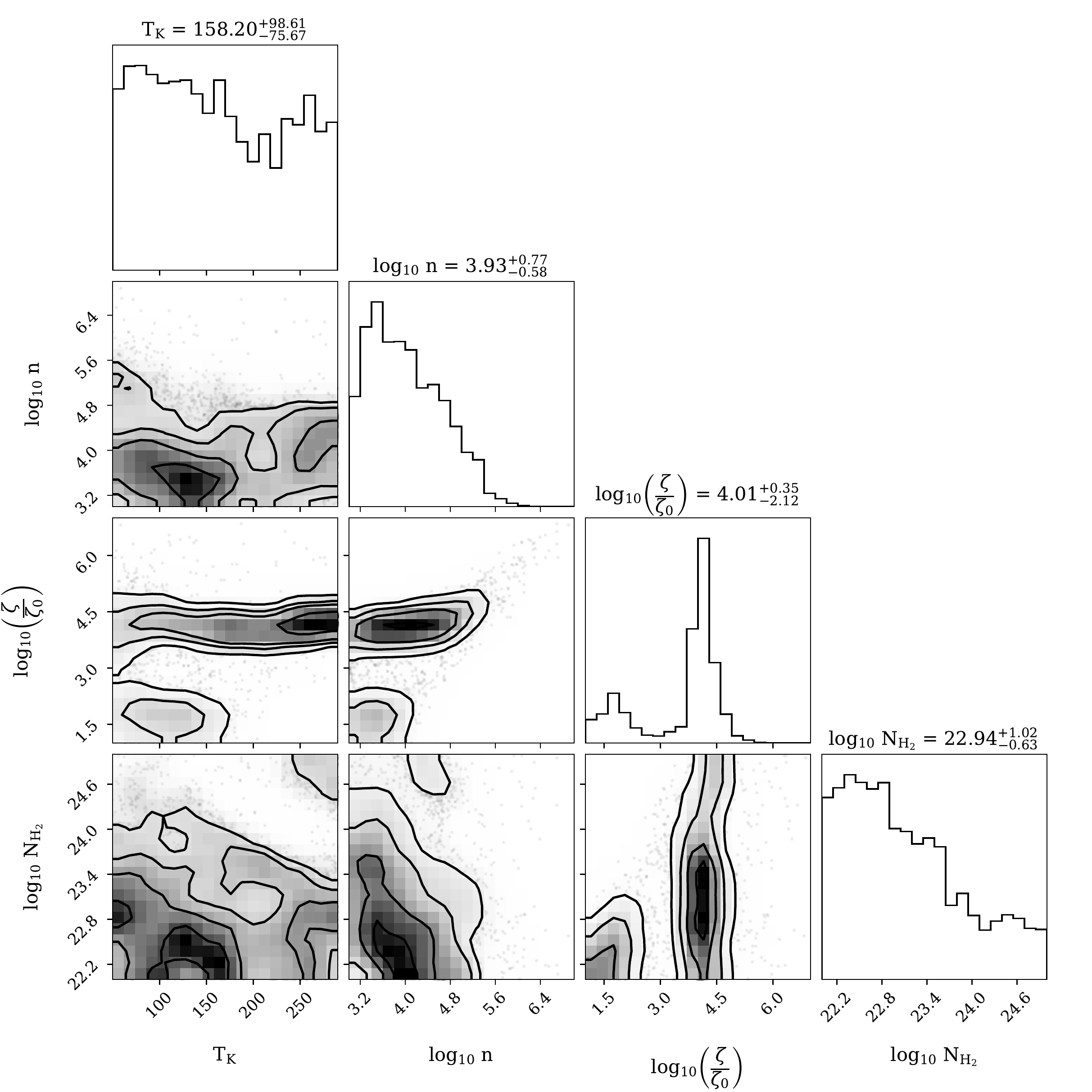}
    \caption{Modeling results for GMC 10.}
    \label{fig:corner10}
\end{figure*}

\begin{figure*}[htbp]
    \centering
    \includegraphics[trim= 4mm 5mm 0mm 0mm, scale=0.55]{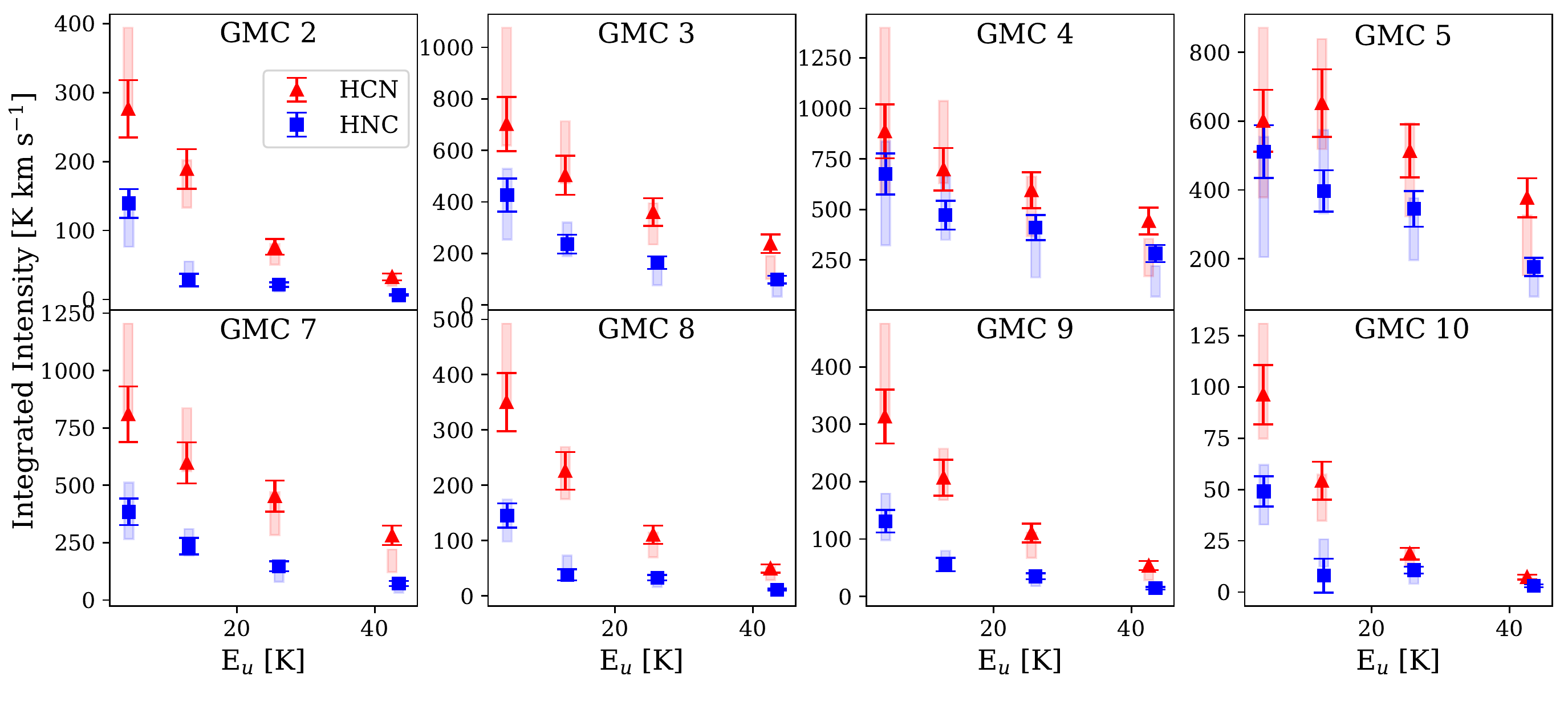}
    \caption{Observed (triangles and squares) versus modeled (shaded bars) flux for HCN (red) and HNC (blue). Observed error bars indicate the 1$\sigma$ uncertainty range. Shaded rectangles show the inner 67\% ($\sim$16th---84th percentile) of our modeled flux distributions.}
    \label{fig:all_fluxes}
\end{figure*}


%\email{eb7he@virginia.edu, %jmangum@nrao.edu,
%jrh@star.ucl.ac.uk,
%nanase.harada@nao.ac.jp, %smartin@eso.org, %nakanisi.k@nao.ac.jp, %mullers@chalmers.se,  %ksakamoto@asiaa.sinica.edu.tw, %yyoshimura@ioa.s.u-tokyo.ac.jp, %nakajima@isee.nagoya-u.ac.jp}
\end{CJK*}
\end{document}